\documentstyle[aps,preprint,prb,graphicx,url]{revtex}
\begin{document}
\newcommand{\etal}{{\em et al.}\/}
\newcommand{\IP}{inner polarization}
\newcommand{\IPF}{\IP\ function}
\newcommand{\IPFs}{\IP\ functions}
\newcommand{\auth}[2]{#2, #1;}
\newcommand{\jcite}[4]{{\it #1} {\bf #4}, {\it #2}, #3}
\newcommand{\et}{}
\newcommand{\twoauth}[4]{#2, #1; #4, #3}
\newcommand{\oneauth}[2]{#2, #1;}
\newcommand{\andauth}[2]{#2, #1;}
\newcommand{\book}[4]{{\it #1} (#2, #3, #4)}
\newcommand{\inbook}[5]{In {\it #1}; #2; #3: #4, #5}
\newcommand{\JCP}[3]{\jcite{J. Chem. Phys.}{#1}{#2}{#3}}
\newcommand{\jms}[3]{\jcite{J. Mol. Spectrosc.}{#1}{#2}{#3}}
\newcommand{\jmsp}[3]{\jcite{J. Mol. Spectrosc.}{#1}{#2}{#3}}
\newcommand{\jmstr}[3]{\jcite{J. Mol. Struct.}{#1}{#2}{#3}}
\newcommand{\cpl}[3]{\jcite{Chem. Phys. Lett.}{#1}{#2}{#3}}
\newcommand{\cp}[3]{\jcite{Chem. Phys.}{#1}{#2}{#3}}
\newcommand{\pr}[3]{\jcite{Phys. Rev.}{#1}{#2}{#3}}
\newcommand{\jpc}[3]{\jcite{J. Phys. Chem.}{#1}{#2}{#3}}
\newcommand{\jpcA}[3]{\jcite{J. Phys. Chem. A}{#1}{#2}{#3}}
\newcommand{\jpca}[3]{\jcite{J. Phys. Chem. A}{#1}{#2}{#3}}
\newcommand{\jpcB}[3]{\jcite{J. Phys. Chem. B}{#1}{#2}{#3}}
\newcommand{\PRA}[3]{\jcite{Phys. Rev. A}{#1}{#2}{#3}}
\newcommand{\PRB}[3]{\jcite{Phys. Rev. B}{#1}{#2}{#3}}
\newcommand{\jcc}[3]{\jcite{J. Comput. Chem.}{#1}{#2}{#3}}
\newcommand{\molphys}[3]{\jcite{Mol. Phys.}{#1}{#2}{#3}}
\newcommand{\mph}[3]{\jcite{Mol. Phys.}{#1}{#2}{#3}}
\newcommand{\APJ}[3]{\jcite{Astrophys. J.}{#1}{#2}{#3}}
\newcommand{\cpc}[3]{\jcite{Comput. Phys. Commun.}{#1}{#2}{#3}}
\newcommand{\jcsfii}[3]{\jcite{J. Chem. Soc. Faraday Trans. II}{#1}{#2}{#3}}
\newcommand{\prsa}[3]{\jcite{Proc. Royal Soc. A}{#1}{#2}{#3}}
\newcommand{\jacs}[3]{\jcite{J. Am. Chem. Soc.}{#1}{#2}{#3}}
\newcommand{\joptsa}[3]{\jcite{J. Opt. Soc. Am.}{#1}{#2}{#3}}
\newcommand{\cjc}[3]{\jcite{Can. J. Chem.}{#1}{#2}{#3}}
\newcommand{\ijqcs}[3]{\jcite{Int. J. Quantum Chem. Symp.}{#1}{#2}{#3}}
\newcommand{\ijqc}[3]{\jcite{Int. J. Quantum Chem.}{#1}{#2}{#3}}
\newcommand{\spa}[3]{\jcite{Spectrochim. Acta A}{#1}{#2}{#3}}
\newcommand{\tca}[3]{\jcite{Theor. Chem. Acc.}{#1}{#2}{#3}}
\newcommand{\tcaold}[3]{\jcite{Theor. Chim. Acta}{#1}{#2}{#3}}
\newcommand{\jpcrd}[3]{\jcite{J. Phys. Chem. Ref. Data}{#1}{#2}{#3}}
\newcommand{\science}[3]{\jcite{Science}{#1}{#2}{#3}}
\newcommand{\CR}[3]{\jcite{Chem. Rev.}{#1}{#2}{#3}}
\newcommand{\bbpc}[3]{\jcite{Ber. Bunsenges. Phys. Chem.}{#1}{#2}{#3}}

\draft
\title{Benchmark {\em ab initio} energy profiles for the gas-phase
S$_N$2 reactions Y$^-$ + CH$_3$X $\rightarrow$ CH$_3$Y + X$^-$
(X,Y = F,Cl,Br). Validation of hybrid DFT methods}

\author{Srinivasan Parthiban, Gl\^{e}nisson de Oliveira\thanks{Present address: 
Chemistry Department, Pensacola Christian College, 250 Brent Lane, Pensacola, FL 32503}, and 
Jan M.L. Martin\thanks{Author to whom correspondence should be addressed. Email: {\tt comartin@wicc.weizmann.ac.il}}}
\address{Department of Organic Chemistry,
Kimmelman Building, Room 262,
Weizmann Institute of Science,
IL-76100 Re\d{h}ovot, Israel. 
}
\date{{\em J. Phys. Chem. A} manuscript JP0031000; revised October 31, 2000}
\maketitle
\begin{abstract}
The energetics of the gas-phase S$_N$2 reactions 
Y$^-$ + CH$_3$X $\longrightarrow$ CH$_3$Y + X$^-$
(where X,Y = F, Cl, Br), were studied
using (variants on) the recent W1 and W2 {\em ab initio} computational
thermochemistry methods. These calculations involve CCSD and CCSD(T)
coupled cluster methods, basis sets of up to $spdfgh$ quality, 
extrapolations to the one-particle basis set limit, and contributions
of inner-shell correlation, scalar relativistic effects, and
(where relevant) first-order spin-orbit coupling. Our computational
predictions are in excellent agreement with experimental data 
where these have small error bars; in a number of other instances
re-examination of the experimental data may be in order.
Our computed benchmark data (including cases for which experimental
data are unavailable altogether) are used to assess the quality
of a number of semiempirical compound thermochemistry schemes such
as G2 theory, G3 theory, and CBS-QB3, as well as a variety of density 
functional theory methods. Upon applying some modifications to the 
level of theory used for the reference geometry (adding diffuse
functions, replacing B3LYP by the very recently proposed mPW1K 
functional [Lynch, B.J.; Fast, P.L.; Harris, M.; Truhlar, D.G. 
\jpca{104}{4811}{2000}]),
the compound methods appear to perform well. Only the 'half-and-half'
functionals BH\&HLYP and mPWH\&HPW91, and the empirical mPW1K
functional, consistently find all required stationary points;
the other functionals fail to find a transition
state in the F/Br case.
BH\&HLYP and mPWH\&HPW91 
somewhat overcorrect for the tendency of B3LYP (and, to a somewhat
lesser extent, mPW1PW91) to underestimate barrier heights. 
The Becke97 and Becke97-1 functionals perform similarly to B3LYP
for the problem under study, 
while the HCTH and HCTH-120 functionals both appear 
to underestimate central barriers. HCTH underestimates complexation
energies; this problem is resolved in HCTH-120.
mPW1K appears to exhibit the best
performance of the functionals considered, although its energetics
are still inferior to the compound thermochemistry methods. mPW1K,
however, appears to be very suitable for generating reference 
geometries for more elaborate thermochemical methods in kinetics
applications.
\end{abstract}
\newpage
\section{Introduction}

Due to the central importance of bimolecular nucleophilic substitution
(S$_N$2) reactions in organic chemistry,\cite{shaik92,ingold69} the
prototype S$_N$2 reactions
\begin{equation}
{\rm Y}^- + {\rm CH}_3{\rm X} \longrightarrow {\rm CH}_3{\rm Y} + {\rm X}^- ~~~~~~ {\rm (X,Y= F,Cl,~and~Br)}
\end{equation}
have aroused considerable interest in the past three decades. 
(Halomethanes have also received considerable attention in the area
of atmospheric chemistry in connection with global warming\cite{ghg91}
and ozone layer destruction.\cite{wmo98})
Both theoretical and experimental studies (see Refs. 
\cite{hase94,brauman98} for reviews) indicate that the preferred 
gas phase reaction pathway involves a backside attack of the halide ion, 
Y$^-$, at the carbon atom followed by the familiar `Walden inversion'
of the CH$_3$ group. The resulting reaction profile (Figure 1) exhibits 
two local minima, i.e. entry and exit channel ion-molecule complexes 
Y$^-$$\cdots$CH$_3$X and YCH$_3$$\cdots$X$^-$, connected by a central 
transition state [Y$\cdots$CH$_3$$\cdots$X]$^-$, which has $D_{3h}$ and
$C_{3v}$ symmetries in the identity (X=Y) and nonidentity (X$\neq$Y)
cases, respectively.  Although the qualitative form of this reaction pathway 
is widely accepted for substitution reactions in the gas phase, there 
is still considerable uncertainty about the exact energetics.

B{\"o}hme \etal\cite{bohme70} and Brauman \etal\cite{brauman79} 
were the first to investigate the gas-phase S$_N$2 reactions experimentally.
Brauman and coworkers concluded that the measurements were best explained
by a double-well potential with a central barrier. Subsequent experimental
studies\cite{caldwell84,bierbaum88,li96,deturi97}
for a series of  anionic nucleophiles 
with alkyl halides revealed that changes in the
nucleophile, leaving group, and alkyl moiety leads to a wide 
variation of reaction rate constants; their observed variation 
was attributed to the central barrier height.

The double-well S$_N$2 potential energy surface also finds abundant theoretical
support from {\em ab initio} calculations, which are currently one of the most useful
tools for evaluating reaction potential energy profiles. 
For instance, Chandrasekhar \etal\cite{chandra85} presented a 
comprehensive examination of Cl/Cl identity S$_N$2 reaction at 
the HF/6-31G* level, and Tucker and Truhlar\cite{tucker89} examined the S$_N$2 reactions
at the MP2/6-31G* level.  Wladkowski et al.\cite{Wla94} studied the 
F/F identity S$_N$2 reaction by large-scale coupled cluster 
theory involving single and double excitation operators with an {\em a posteriori} 
a quasiperturbative 
treatment of the effects of connected triple substitutions (CCSD(T)) 
(see Refs. \cite{lee,bartlett} for reviews).  Later, Radom, Pross and coworkers have
carried out {\em ab initio} molecular orbital calculations at the G2(+)
level of theory for the back-side identity\cite{radomid95} and 
nonidentity\cite{radomnonid96} S$_N$2 reactions.
These authors have also investigated the identity front-side S$_N$2 reactions with retention
of configuration.\cite{radomret96} The G2(+) theory is essentially G2 theory
carried out from MP2/6-31+G* (rather than MP2/6-31G*) geometries and employing
scaled HF/6-31+G* (rather than HF/6-31G*) zero-point energies. For the bromine
and iodine containing systems, these authors employed Hay-Wadt\cite{hay85}
relativistic effective core potentials (RECPs).  Botschwina and 
coworkers examined the stationary points of the potential surface 
for the F/Cl nonidentity S$_N$2 reaction\cite{botsch97} and for the Cl/Cl identity 
S$_N$2 reaction\cite{botsch98} by means of large-scale CCSD(T) calculations.
Finally, a referee brought a very recent large-scale coupled cluster study 
by Schmatz et al.\cite{Sch2000} to our attention.

Despite the well known successes (e.g.\cite{dftstudies,martin95}) of the
increasingly popular DFT (density functional theory) methods,\cite{dft}
their performance for transition state structures and reaction barrier heights leaves
something to be desired.  For instance, Durant\cite{durant96} 
found that the B3LYP, B3P86 and B3PW91 functionals all systematically 
underestimated barrier heights, while only the Becke half-and-half/Lee, 
Yang and Parr (BH\&HLYP)\cite{BHLYP} functional predicted transition state barrier 
heights reasonably well --- despite the fact that its performance for thermochemical 
and other properties is generally substantially poorer than that of 
B3LYP and B3PW91. Baker \etal\cite{baker95} arrived at a similar conclusion
stating that the currently available 
density functionals are unable to provide a correct description of the 
transition states. For the prototype S$_N$2 reactions
(Cl/Cl and Cl/Br) considered here, Radom and coworkers\cite{radomdft96}
found that the popular B3LYP\cite{Bec93,LYP} exchange-correlation functional 
significantly underestimated the
overall and central barrier heights compared to the G2(+) and experimental results.

Nevertheless, the size of the
systems involved in kinetic and mechanistic problems of organic and organometallic 
interest often makes DFT the only practical option. As a matter of fact,
our group has recently reported DFT studies of the mechanism 
of competitive intramolecular C-C and C-H bond activation in 
rhodium(I) pincer complexes\cite{andreas} and of the Heck reaction.\cite{heck}

Aside from BH\&HLYP, better performance for barrier heights has been claimed 
for a number of newer exchange-correlation functionals.  For
example, Adamo and Barone\cite{mpw1pw98} found that their mPW1PW91
(modified Perdew-Wang 1991 1-parameter hybrid exchange with
Perdew-Wang 1991 correlation\cite{pw91}) at least correctly predicts a positive
overall barrier for  the  Cl/Cl identity S$_N$2 reaction, although
it is still being underestimated. 
Very recently, 
Truhlar and coworkers\cite{mpw1k} proposed a new hybrid model called the 
modified Perdew-Wang 1-parameter model for kinetics (mPW1K). In this empirical 
functional, the coefficient $X$ for admixture of "exact" Hartree-Fock exchange
\begin{equation}
V_{XC} = X V_{X,HF} + (1-X) V_{X,mPW1} + V_{C,PW91}
\end{equation}
(where $X$=1/4 for standard mPW1PW91)
was determined (using the fairly small 6-31+G* basis set) 
by minimizing the average deviation from a set of 40 barrier
heights (20 forward, 20 reverse) obtained from a combination of experiment
and theory (see Ref.\cite{mpw1k} for details). (Note that the Walden inversion, 
or for that matter cationic or anionic reactions of any kind, were not part of 
the parametrization set.) 
It was found that mPW1K reduced the mean unsigned error
in reaction barrier heights by a factor of 2.4 over mPW1PW91 and
by a factor of 3 over B3LYP. 

Theoretical models such as Transition State Theory (TST)\cite{tst} and
Rice-Ramsperger-Kassel-Marcus (RRKM)\cite{rrkm} theory were also employed
to examine the S$_N$2 reactions. Results from such studies
(see \cite{rrkmhase94} and references therein) suggested that the assumption
of statistical behavior in ion-molecule intermediate complexes is not valid.
This "nonstatistical" behavior has been documented for several halide-methyl
halide reactions and a thorough discussion is given by Hase.\cite{hase94} 
Classical trajectory simulations performed by
Hase and coworkers\cite{traj98} questioned the basic
assumptions of statistical theories and found that the trajectory
calculations are very useful in  interpreting the kinetics and dynamics
of S$_N$2 reactions.

Despite the enormous amount of work in the past, there are still significant gaps in the
experimental data for the gas-phase S$_N$2 reactions and, even where
data are available, the results often possess large uncertainties.
Recently, two computational thermochemistry methods known as W1 and W2
(Weizmann-1 and Weizmann-2) theory\cite{w1w2} have been developed in our
laboratory. These are free of parameters derived from experiment and
on average can claim `benchmark accuracy' (defined
in Ref.\cite{w1w2} as a mean absolute error of 1 kJ/mol, or 0.25 kcal/mol)
for molecular
total atomization energies (TAEs) of first-and second-row compounds.
The primary objective of the present study is to obtain high-quality
energetic data for reaction (1) by means of W1 and W2 theory. Using these
benchmark data, we shall then 
examine the performance of various DFT methods and {\em ab initio} 
computational thermochemistry methods such as G1,\cite{g1paper} G2,\cite{g2paper} 
G3,\cite{g3paper} and CBS-QB3\cite{cbs-qb3} theories.

\section{Computational methods}

All calculations were carried out on the 4-processor Compaq ES40 of
our research group, and on the 12-CPU SGI Origin 2000 of the Faculty of
Chemistry. 

Energetics for the gas-phase stationary points for all six surfaces
(i.e. F/F, Cl/Cl, Br/Br, F/Cl, F/Br, and Cl/Br) were obtained by
means of the W1$'$ method described in Refs.\cite{w1w2,so3}. The
W1$'$ method\cite{so3} is a minor variation on W1 theory\cite{w1w2}
that exhibits improved accuracy for second-row systems at no
additional computational cost. For a detailed description and 
theoretical and empirical arguments for each step, see
Ref.\cite{w1w2}; we shall merely summarize the main points here
for the sake of clarity. The basis sets employed are mostly Dunning's
augmented correlation consistent $n$-tuple zeta\cite{Dun89,Ken92,Dun97}
(aug-cc-pV$n$Z) basis sets; for
second-row atoms high-exponent $d$ and $f$ functions were added (denoted
'+2d' or '+2d1f') as recommended in Ref.\cite{so2} for accommodating 
inner polarization. Since the standard aug-cc-pV$n$Z basis sets for 
bromine\cite{Wil99} already contain quite high-exponent $d$ functions in
order to describe the $3d$ orbitals, no 'inner polarization' functions
were deemed to be necessary on Br. We may distinguish the following six
components in the `bottom-of-the-well' TAE 
at the W1$'$ level:
\begin{itemize}
\item The SCF component of the TAE 
is obtained using the aug-cc-pVDZ+2d, aug-cc-pVTZ+2d, and aug-cc-pVQZ+2d1f
basis sets, and extrapolated to the infinite-basis limit using the
geometric expression\cite{Fel92} 
$A+B\cdot C^{-L}$, where the `cardinal number' L=\{2,3,4\} for these three
basis sets. (It is identical to the maximum angular momentum 
present for nonhydrogen atoms. Regular cc-pV$n$Z basis sets
were used on hydrogen atoms throughout.)
\item The CCSD (coupled cluster with all singles and
doubles\cite{Pur82}) valence correlation contribution to TAE 
is obtained using the aug-cc-pVTZ+2d and aug-cc-pVQZ+2d1f
basis sets, then extrapolated to the infinite basis limit using
the expression $A+B/L^{3.22}$. 
\item The (T) connected triple excitations component\cite{Rag89} of TAE
was computed using the aug-cc-pVDZ+2d and aug-cc-pVTZ+2d basis sets,
and extrapolated to the infinite basis limit using
the expression $A+B/L^{3.22}$.
\item The inner-shell correlation contribution was computed as 
the difference between CCSD(T)/MTsmall\cite{w1w2} values with and without
constraining the inner-shell orbitals to be doubly occupied. The very 
deep-lying chlorine (1s) and bromine (1s,2s,2p) orbitals were doubly 
occupied throughout; the `inner-shell correlation' thus represents
carbon (1s), chlorine (2s,2p) and bromine (3s,3p,3d) correlation. 
(Basis set superposition error, BSSE, can be an issue for inner-shell correlation
energies in heavier element systems;\cite{Bau98} our experience\cite{w1w2}
suggests that BSSE in the W1/W2 inner shell correlation contributions 
largely cancels with basis set incompleteness.)
\item The scalar relativistic contribution was computed as expectation
values of the one-electron Darwin and mass-velocity (DMV)
operators\cite{Cow76,Mar83} for the ACPF/MTsmall (averaged coupled
pair functional\cite{Gda88}) wave function, with all inner-shell
electrons correlated except for chlorine (1s) and bromine (1s,2s,2p).
\item The spin-orbit contribution to TAE, in the present case of
all-closed-shell systems, is nothing more than the sum of the atomic
fine structure corrections.
\end{itemize}

Where our computational hardware permitted (in practice, 
for F/F, Cl/Cl, and the Br/Br transition state), 
we also carried out even more demanding
W2h calculations. In W2 theory, the same steps occur as above,
except that the three valence basis sets are aug-cc-pVTZ+2d1f,
aug-cc-pVQZ+2d1f, and aug-cc-pV5Z+2d1f (with L=3,4, and 5, respectively)
and that the extrapolation formula\cite{Hal98} used for the CCSD and (T) 
steps is simply $A+B/L^3$. The W2h variant\cite{w2h} indicates, in
this particular case, the use of unaugmented cc-pV$n$Z basis sets on carbon.
The largest basis set CCSD step was carried out using the direct algorithm
of Lindh, Sch\"utz, and Werner.\cite{dirccsd}
All these calculations were performed using MOLPRO 98.1\cite{molpro98}
and a driver for the W1/W2 calculations\cite{autoW1W2} written in
MOLPRO's scripting language. Reference geometries were obtained 
primarily using the B3LYP\cite{Bec93,LYP} density functional method, 
which employs the Lee-Yang-Parr\cite{LYP}
correlation functional in conjugation with a hybrid exchange
functional first proposed by Becke.\cite{Bec93}

A number of lower-level procedures were validated against the W1$'$ and
W2h results. These include the following set of DFT exchange-correlation
functionals and basis sets: B3LYP/cc-pVTZ(+X), BH\&HLYP/cc-pVTZ(+X), mPW1PW91/cc-pVTZ(+X),
mPW1K/6-31+G*, mPW1K/cc-pVDZ(+X), mPW1K/cc-pVTZ(+X), where
(+X) indicates that diffuse functions are included only for halogens.
BH\&HLYP\cite{BHLYP} is essentially the B3LYP method, with the exception that
the fraction of HF exchange is 50\% (H\&H denotes "half and half").
Analogous to BH\&HLYP, we have also performed mPWH\&HPW91.\cite{mpw1handhpw91}
In addition, we carried out calculations using the standard G1, G2, G3, and
CBS-QB3 model chemistries. The MP2/6-31G* and B3LYP/6-311G(2d,d,p)
levels of theory (used for the reference geometries in G3\cite{g3paper} and CBS-QB3,\cite{cbs-qb3}
respectively) fail to find stationary points for several of the
ion-molecule complexes in the nonidentity cases (because of the absence
of diffuse functions). Therefore, we have defined, by analogy with Radom
and coworkers,\cite{radomid95,radomnonid96,radomret96} G3(+) and CBS-QB3(+) model chemistries where 
MP2/6-31+G* and B3LYP/6-311+G(2d,d,p) reference geometries, respectively,
are used. The G2(+) results quoted in the tables are taken from 
Radom and coworkers.\cite{radomid95,radomnonid96} All of these calculations were
carried out using Gaussian 98 rev. A7\cite{g98revA7} or trivial
modifications thereof. Following the recommendations in Ref.\cite{grids},
larger grids than the default were used in the DFT calculations
if necessary, specifically a pruned (99,590) grid for integration and 
gradients, and a pruned (50,194) grid for the solution of the coupled
perturbed Kohn-Sham equations. 

Finally, following very recent suggestions in the literature\cite{HCTHforTS} that some of these functionals may 
perform better for transition states,
some calculations using the novel B97 (Becke-1997),\cite{Becke97}
B97-1 (reparametrized Becke-1997),\cite{hcth} HCTH (Hamprecht-Cohen-Tozer-Handy),\cite{hcth} and HCTH-120 (reparametrization of HCTH including anions and
weakly interacting systems)\cite{hcth120}
functionals were carried out by means of 
a slightly modified version of NWCHEM 3.3.1.\cite{nwchem}

\section{Results and discussion}

\subsection{Reference Geometries}

Reference geometries for the W1$'$ calculations were mostly obtained at
the B3LYP/cc-pVTZ+1 level, where the `+1' signifies the addition of 
a high-exponent $d$ function on second-row elements.\cite{sio} (The Br basis set already
includes high-exponent $d$ functions to cover the $(3d)$ orbital.)
It was previously shown\cite{martin95} that B3LYP/cc-pVTZ geometries for 
stable molecules are generally within a few thousandths of an \AA\ from experiment,
as well that the use of B3LYP/cc-pVTZ+1 rather than much costlier CCSD(T)/cc-pVQZ+1
reference geometries insignificantly affects computed energies.\cite{w1w2} 
(Modifications of popular computational thermochemistry methods
that use DFT reference geometries include variants\cite{morokuma95,charlie95}
of G2 theory, G3//B3LYP,\cite{g3b3} and CBS-QB3.\cite{cbs-qb3}) In some
cases where B3LYP fails to locate the required stationary point, we used
mPW1K/cc-pVTZ(+X) reference geometries. For the W2h calculations, 
CCSD(T)/cc-pVQZ+1 reference geometries were used. The geometries of
all the structures involved in the present study calculated at various
levels of theory  are provided in the Supplementary Material. 

\subsection{Energetics}

In order to assess the accuracy of W1$'$ and W2h results, we consider first
the total atomization energies (TAEs) of CH$_3$X and electron affinities (EAs) of X$^-$.
A summary of our computed results and their components 
for the reactants/products of the S$_N$2 reactions is presented in Table 1. 
The final energies presented 
in the last column of the Table correspond to EAs of X 
(X = F, Cl and Br) and TAEs without zero-point 
vibrational energy (ZPVE) of CH$_3$X. The 
inner-shell correlation contributions are all positive and the largest is 
1.49 kcal/mol for CH$_3$Br. 
The core correlation contribution for TAE(CH$_3$X) is found to increase in 
the order F $<$ Cl $<$ Br, while for EA(X) it increases in 
the order Cl $<$ F $<$ Br. The importance of Darwin and 
mass-velocity corrections increases, as expected, with increasing 
atomic number (Z) of X and its contribution becomes substantial 
when X = Br.

It is perhaps more pertinent for our purposes to examine the relative 
energies (with respect to reactants)
of the ion-molecule complexes and 
transition state structures. Table 2 presents  
W1$'$ and W2h results for identity reactions and only W1$'$ results 
for nonidentity reactions. W2h calculations for the nonidentity 
reactions are extremely expensive as the reaction intermediates are 
less symmetric. Moreover, the size of the bromine atoms prevents us from
performing a W2h calculation on the identity Br$^-$$\cdots$CH$_3$Br 
ion-molecule complex. Likewise, we could not obtain 
the core correlation contributions for the Br$^-$$\cdots$CH$_3$Br 
ion-molecule complex at the W1$'$ level of theory. It was previously
established\cite{martintruhlar} that the inclusion of connected triple 
excitations in CCSD(T) is absolutely necessary for reliable core
correlation contributions: the $n^3N^4$ CPU time dependence of the (T) step
dominates the required CPU time for Cl and Br. Both the size of the
halogen atoms and the reduced symmetry prevented us from 
performing core-correlation calculations
for nonidentity S$_N$2 reactions, except for the F/Cl nonidentity case.  

From Table 2, it can be seen that the final W1$'$ and W2h energy values 
for the identity reactions are very close to each other. Considering the very close 
agreement between W1$'$ and W2h results, the conclusion is warranted 
that the results from W1$'$ theory can be used as reference values to 
compare the results from other methods when W2h results are not available. 
As a general observation, the core contributions for the transition state 
structures are noticeably larger than for the ion-molecule complexes (see Table 2). 
Although the core-correlation contribution is small in absolute terms, 
its relative contribution to the likewise small overall barrier heights can
be substantial.  For example, it is 0.36 kcal/mol for 
$\lbrack$F$\cdots$CH$_3$$\cdots$F$\rbrack$$^-$, while the 
total energy is -0.37 kcal/mol. Likewise, the core correlation 
contributions for $\lbrack$Cl$\cdots$CH$_3$$\cdots$Cl$\rbrack$$^-$ and
$\lbrack$Br$\cdots$CH$_3$$\cdots$Br$\rbrack$$^-$ are nearly 10\% and 25\%
of the overall barrier (relative to reactants), respectively. At the W2h 
level, the core correlation contribution
to the total energy increases slightly. Scalar relativistic effects
exhibit similar trends as those of core correlation, but the
effects are fairly small. Only Cl$^-$$\cdots$CH$_3$Br,
$\lbrack$Cl$\cdots$CH$_3$$\cdots$Br$\rbrack$$^-$ and
Br$^-$$\cdots$CH$_3$Cl exhibit noticeable scalar relativistic 
contributions, due to the presence of the heavy halogen Br. 

Among the identity reactions 
only $\lbrack$F$\cdots$CH$_3$$\cdots$F$\rbrack$$^-$
has a transition state below the reactants energy level.
In the  nonidentity case all the transition state structures lie below the
reactants. 

The computed final heats of formation ($\Delta H^\circ_f$) of 
CH$_3$X 
in kcal/mol are compared with experiment in Table 3. 
Both W1$'$ and W2h values are presented after accounting for ZPVEs and 
thermal corrections calculated at the B3LYP/cc-pVTZ+1 level. At this level 
the ZPVEs, after scaling by 0.985,\cite{w1w2} are found to be 24.16, 23.26 and 
22.89 kcal/mol, respectively, for CH$_3$F, CH$_3$Cl and CH$_3$Br. 
The corresponding thermal corrections are -1.92, -1.89 and 
-3.68 kcal/mol. 
The computed $\Delta H^\circ_f$ 
value for CH$_3$Cl lies within the experimental error bar:
the experimental value for CH$_3$F is a crude estimate ($\pm$7 kcal/mol)
and our computed value is certainly more reliable. Our calculated value for
CH$_3$Br is slightly outside the experimental error bar: some of the
discrepancy could be due to the limitations of the scalar relativistic
treatment. As shown by Bauschlicher,\cite{charlie99} the simple DMV
correction starts to exhibit minor deficiencies for third-row compounds;
for first- and second-row compounds, it is in excellent agreement with more
rigorous treatments.\cite{charlie2000,martin99}

Also included in Table 3 are calculated electron affinities of
X (X = F, Cl and Br) in eV together with experimental results.
Using a similar approach, but with even larger $spdfghi$ basis sets
as well as full CI corrections, we were able\cite{glen99} to reproduce
the experimental EAs of the 1st- and 2nd-row atoms to within $\pm$0.001 eV
on average.
The presently calculated W2h 
results of F and Cl EAs differ by only about 0.001 eV from these 
benchmark values (EA(F)=3.403 eV and EA(Cl)=3.611 eV), and the W2h results 
for F, Cl and Br are all within 0.003 eV of experiment. Although the W1$'$ 
values differ about 0.01 eV for F and Cl and 0.02 eV for Br, this is 
comparable to the W1/W2 target accuracy (0.25 kcal/mol on average).
The performance of the W1$'$ and W2h methods for the
reactants and products is obviously encouraging for the study of the
problem at hand.

\subsection{S$_N$2 Reactions}

The reaction mechanism with the double-well potential energy surface
for the gas-phase S$_N$2 reactions is shown in Figure 1.
Obviously, the energy profile is symmetric for the identity
reactions (Figure 1a), and asymmetric for the
nonidentity reactions (Figure 1b). The complexation energy ($\Delta$$H$$_{comp}$),
central barrier ($\Delta${\it H}$^\ddag$$_{cent}$), and overall activation
barrier relative to the separated reactants 
($\Delta${\it H}$^\ddag$$_{ovr}$) are defined in Figure 1.
In
the nonidentity case, the following additional quantities are
defined in Figure 1b: overall enthalpy change for the reaction
($\Delta${\it H}$_{ovr}$) and the central enthalpy difference $\Delta${\it H}$_{cent}$
between product and reactant ion-molecule complexes, {\bf 3} and {\bf 1}.

\subsection{Identity Reactions}

Complexation energies ($\Delta$$H$$_{comp}$), overall barrier heights 
($\Delta${\it H}$^\ddag$$_{ovr}$) and
central barriers ($\Delta${\it H}$^\ddag$$_{cent}$) obtained from W1$'$ 
and W2h methods are compared in Table 4 with DFT, Gn, and CBS-QB3 methods together with
available experimental values.

It should be emphasized that the experimental data for the S$_N$2 
reactions are insufficient and the available data are subject to large
uncertainties. Therefore, it would be appropriate to analyze the 
performance of various methods with respect to Wn methods. 
First of all, note that the mPW1K/6-31+G* $\Delta$$H$$_{comp}$ 
(13.55 kcal/mol) for the F/F case is very close to the W1$'$ and
W2h results (13.66 and 13.72 kcal/mol, respectively). 
mPW1K/cc-pVDZ(+X) and mPW1K/cc-pVTZ(+X) methods however predict
lower $\Delta$$H$$_{comp}$ values. In fact, the B3LYP, B97, HCTH-120, mPW1PW91, 
mPWH\&HPW91 and mPW1K methods 
all predict roughly 1 kcal/mol lower complexation energies, while BH\&HLYP and B97-1
agree well with W1$'$ and W2h. (The HCTH $\Delta$$H$$_{comp}$ is much lower than the others,
{\em vide infra}.) G2 and CBS-QB3 values are close to the Wn results while the 
G3 method predicts higher complexation energy compared to the Wn methods. 
Inclusion of diffuse functions for the Gn and CBS-QB3 
reference geometries (i.e. Gn(+) and CBS-QB3(+)) increases the 
$\Delta$$H$$_{comp}$ value by 0.4--0.6 kcal/mol. 

A comparison of overall barrier heights is presented in the third column
of Table 4. Both W1$'$ (-0.37 kcal/mol) and W2h 
(-0.34 kcal/mol) theories predict negative barrier heights in the F/F case
and the values are very close. The CBS-QB3(+) result is in excellent agreement
therewith; all G$n$ theories predict barrier heights that are lower by 
1 kcal/mol, with further lowering seen at the G2(+) and G3(+) levels. 
Among the DFT methods considered, only
mPW1K/6-31+G* and HCTH/cc-pVDZ(+X) fortuitously predict overall barrier heights 
close to the Wn results: basis set extension for mPW1K leads to
positive overall barrier heights, which are likewise found for the 
"half and half" functionals. B3LYP, B97(-1) and HCTH-120 all 
significantly underestimate the barrier, mPW1PW91 to a lesser extent.

For the chlorine identity gas-phase S$_N$2 reactions, fairly accurate
experimental values are available and are presented in 
Table 4. The experimental values reported by Li and coworkers\cite{li96} 
correspond to the standard state.  Hence, thermal corrections and ZPVEs 
are subtracted from experimental values in order to compare with 
the "bottom of the well" calculated values. It is noteworthy
that the W1$'$ (10.54 kcal/mol) and W2h (10.94 kcal/mol) 
complexation energies are in good agreement with the 
experimental value (10.53 kcal/mol). CBS-QB3 results are also 
in agreement with the Wn and experimental values, while those
from DFT calculations are less satisfactory as they are about
1 kcal/mol lower. Also note that the G1 and G2 methods 
reproduce the complexation energy well, while G3 results are 0.5 kcal/mol higher 
than the Wn and experimental values.

The overall barrier height for the Cl/Cl reaction is found to be 3.07 and 2.67 kcal/mol
at the W1$'$ and W2h levels of theory. Note first that the experimental
value (2.90 kcal/mol) is very close and lies between the  W1$'$ and 
W2h values. The $\Delta${\it H}$^\ddag$$_{ovr}$  value calculated at the mPW1K/6-31+G*, 
and CBS-QB3 levels of theory as well as the G2(+) value by Radom \etal\cite{radomid95} and 
the CCSD(T)/$spdfg$ value by 
Botschwina\cite{botsch98} agree well with the W2h result.
The mPW1K exchange-correlation functional with the cc-pVDZ(+X) and
cc-pVTZ(+X) basis sets predict somewhat higher $\Delta${\it H}$^\ddag$$_{ovr}$ values,
while the G1, G2MP2 and G3(+) values are about 1 kcal/mol lower. B3LYP,
B97, B97-1, and HCTH-120 all predict a  
negative overall barrier for the Cl/Cl system, in disagreement 
with all other methods considered and with experiment. 
BH\&HLYP performs moderately well, while mPWH\&HPW91 predicts a 
larger barrier height (4.50 kcal/mol) than W$n$.
The central barrier values presented in the last column of the Table 4
reveal that the agreement between Wn theories (13.61 kcal/mol) and 
experiment (13.66 kcal/mol) is excellent. 
The G2MP2, G2(+) and CBS-QB3 methods also reproduce the central 
barriers very well. As expected from the overall barrier heights, the DFT results 
are less satisfactory, except for mPW1K/cc-pVDZ(+X) and 
mPW1K/cc-pVTZ(+X) which are in good agreement with the Wn values.

For the bromine identity S$_N$2 reaction, W1$'$ theory predicts 10.03 kcal/mol
for $\Delta$$H$$_{comp}$.
Note that the G2(+) value is in close agreement with
W1$'$ theory. The reported experimental value (11.34$\pm$0.4 kcal/mol) agrees fairly well.
Most DFT levels of theory considered suggest a complexation
energy about 1 kcal/mol lower than the W1$'$ value, except 
mPW1K/6-31+G** which is higher (12.78 kcal/mol, probably an artifact of the
small basis set); B97 and B97-1 which closely bracket the W1$'$ value;
HCTH-120 which is close to the W1$'$ value (see below); and HCTH which is 2.5 kcal/mol
lower than the latter.
The complexation energies for X$^-$$\cdots$CH$_3$X are
found to decrease in the order  F $>$ Cl $>$ Br. This trend was noted
previously by Radom and coworkers,\cite{radomid95} who attributed it
to the electronegativities of the halogens.

The overall 
barrier height for the Br/Br reaction is found to be 1.02 and 0.77 kcal/mol
at the W1$'$ and W2h levels of theory. Of the various exchange-correlation
functionals considered, only BH\&HLYP, mPWH\&HPW91, and mPW1K find positive
barriers (as do the G$n$ theories). It should be pointed out that the DFT results for this
system display appreciable basis set sensitivity: for instance, the
mPW1K/6-31+G* overall barrier has the wrong sign. 
It is interesting to note that 
the complexation energy derived from the experimental overall 
(1.73 kcal/mol\cite{wilbur93}) and central 
(11.68 kcal/mol\cite{pellerite83}) barrier heights is 9.95 kcal/mol 
while the reported experimental complexation energy (11.34 kcal/mol)\cite{li96} is 
inconsistent with the derived value. In fact, the derived value is 
in excellent agreement with the W1$'$ value (10.03 kcal/mol).  
This clearly suggests that the experimental data should be re-examined.

The performance of both B97 and B97-1 for the identity reactions is quite similar
to that of B3LYP. While the `pure DFT' HCTH functional appears to yield markedly
better overall barrier heights, this comes at the expense of significantly
underestimated complexation energies (and severely overestimated ion-molecule distances,
see Supplementary Material). It was previously noted\cite{tuma} that HCTH
severely underestimates interaction energies of H-bonded complexes;
this was ascribed to the absence of anions and H-bonded dimers in the
original HCTH parametrization set. A reparametrization\cite{hcth120} 
against an enlarged sample of high-quality {\em ab initio} energies, denoted 
HCTH-120, eliminates this particular problem.\cite{tuma} For the identity S$_N$2
reactions, we find that complexation energies (and ion-molecule distances)
are dramatically improved
compared to HCTH: no corresponding improvement is however seen for the
central barrier heights, and the overall barrier heights deteriorate accordingly.

Overall, the DFT methods are less satisfactory for barrier height calculations.
Although the performance of mPW1K/6-31+G* method for F/F and Cl/Cl reactions
was excellent, it is not the ultimate low cost method for barrier 
heights as it has predicted a negative barrier for the Br/Br system.
This behavior illustrates the inadequacy of the 6-31+G* basis set for Br:
the more extended correlation consistent basis sets with the mPW1K  
exchange-correlation functional do
predict the sign correctly. In addition, Gn(+) and CBS-QB3(+)
provide an acceptable account of reaction energetics.

\subsection{Nonidentity Reactions}

A comparison of computed and experimental complexation energies for 
the nonidentity S$_N$2 reactions is provided in Table 5. 
For the F$^-$$\cdots$CH$_3$Cl ion-molecule complex
we could find a stationary point neither at the MP2/6-31G* level of theory
used for the G2 and G3 reference geometries
nor at the B3LYP/6-311G(2$d,d,p$) level used for the CBS-QB3
reference geometries; at these levels of theory, the 
optimization leads to Cl$^-$$\cdots$CH$_3$F even if the initial
geometries were chosen to correspond to F$^-$$\cdots$CH$_3$Cl. Addition of diffuse
functions to the basis set for the reference geometry remedies the problem.
Similarly, 
in the F/Br case only the Br$^-\cdots$CH$_3$F complex is found as a 
stationary point at the MP2/6-31G* level of theory, and the transition
state and second ion-molecule complex only appear when diffuse functions
are added to the basis set.
Furthermore, and regardless of the basis set employed,
none of the DFT functionals except mPW1K, mPWH\&HPW91, and BH\&HLYP
find a transition state or a F$^-\cdots$CH$_3$Br complex.
(The CBS-QB3 method is not defined for Br and hence no CBS-QB3 data are 
presented for the F/Br and Cl/Br ion-molecule complexes.) Table 5 also 
presents large-scale CCSD(T) energetics for the F/Cl\cite{botsch97} and
Cl/Br\cite{Sch2000} cases reported by Botschwina and coworkers.
Available experimental values
are presented at the end of the Table with uncertainties in  
parentheses. 

Examination of Table 5 indicates that the $\Delta$$H$$_{comp}$
values strongly depend on the nucleophile (Y$^-$), decreasing
in the order F$^-$ $>$ Cl$^-$ $>$ Br$^-$. They also depend on 
the leaving group (X$^-$), in the order CH$_3$F $<$ CH$_3$Cl $<$ CH$_3$Br.
Similar observations were made earlier by Radom and coworkers.\cite{radomnonid96}

Comparison of the complexation energies obtained from various methods with
W1$'$ theory indicates that all DFT results for Cl$^-$$\cdots$CH$_3$F
are lower by 1 kcal/mol. The only available experimental value\cite{larson84} for 
Cl$^-$$\cdots$CH$_3$F  ($\Delta H^o$ = 11.41 kcal/mol) has an uncertainty 
of 2.01 kcal/mol. Comparison of this value with the calculated values 
suggest that more accurate measurements are in order.
For Cl$^-$$\cdots$CH$_3$Br and Br$^-$$\cdots$CH$_3$Cl, rather
more accurate high-pressure mass spectrometry data
are available (12.54 and 11.01 kcal/mol).
The W1$'$ values (11.91 and 10.32 kcal/mol) are
very close to the experimental results, considering the
experimental uncertainty of 0.4 kcal/mol. While the mPW1K/6-31+G* values
for Cl$^-$$\cdots$CH$_3$Br and Br$^-$$\cdots$CH$_3$Cl are fortuitously
within the experimental error bars, the other DFT methods predict lower values. 
Also note that G2(+) predicts complexation energies close to W1$'$ and 
experiment for Cl$^-$$\cdots$CH$_3$Br, while
the Br$^-$$\cdots$CH$_3$Cl value is small. A complete assessment of CBS-QB3 
is not possible as it could not be applied to the bromine-containing systems.

Like for the identity reactions, complexation energies are 
significantly underestimated (and ion-molecule distances 
overestimated by up to 0.3 \AA: see Supplementary Material) 
by HCTH, and this problem is
mostly remedied by HCTH-120. B97 and especially B97-1
appear to represent an improvement over B3LYP for the
complexation energies.

Calculated overall reaction enthalpies, central enthalpy
differences between reactant and product ion-molecule complexes,
overall barrier heights, and central barrier heights for the
nonidentity S$_N$2 reactions are presented with available
experimental results in Table 6. It needs to be reemphasized that
all values are ``bottom-of-the-well'' (i.e, zero-point exclusive):
that is, the experimental values are presented after subtracting the 
ZPVEs (scaled by 0.985) and thermal corrections obtained using the B3LYP method.

The overall reaction enthalpies of the three nonidentity reactions,
viz F/Cl, F/Br, and Cl/Br, calculated at the W1$'$ level are
-32.65, -41.43 and -8.56 kcal/mol, respectively. The corresponding
experimental values are available and are presented in  Table 6.
The experimental value for the F/Cl reaction (-33.34 kcal/mol) is in close
agreement with the W1$'$ value. B3LYP, B97(-1), mPW1PW91, Gn, and CCSD(T)/$spdfg$ 
results are all in close agreement with the W1$'$ value, but mPW1K, "half and half" and 
CBS-QB3 theories predict 3-5 kcal/mol higher exothermicity.
For the F/Br reaction, the mPW1K/6-31+G* and mPW1PW91/cc-pVTZ(+X)
methods yield overall reaction enthalpies which are quite close to 
the W1$'$ result. The experimental result (-40.20 kcal/mol) is
in good agreement with the best calculated values considering
the uncertainty of 1 kcal/mol. The G$n$ theories 
predict 
exothermicities below, and mPW1K/cc-pV$n$Z(+X) above, the W1$'$ value.
Concerning the Cl/Br nonidentity reaction, the reported experimental
value (-6.86 kcal/mol) differs from the  W1$'$ value by 2 kcal/mol.
As expected, the very recent CCSD(T)/$spdfgh$ results of Botschwina 
and coworkers\cite{Sch2000} are in close agreement with our predictions.
Our results suggest that the Cl/Br experimental data may need to be reconsidered.
Note the significant basis set dependence in the mPW1K results, which
illustrates the inadequacy of the 6-31+G* basis set.

As a general observation, 50:50 admixture of
HF exchange in the DFT theories increases the magnitude 
of the overall reaction enthalpy, and the increase is greater in BH\&HLYP 
than in mPWH\&HPW91.
Performance of B97 and B97-1 for the overall reaction enthalpies is
similar to B3LYP, while HCTH and HCTH-120 represent underestimates
in absolute value.

At the W1$'$ level of theory, the calculated central barrier 
($\Delta${\it H}$^\ddag$$_{cent}$) for the F/Cl system is 2.89 kcal/mol.
G$n$(+), CBS-QB3(+), BH\&HLYP, mPWH\&HPW91, and mPW1K
all reproduce the W1$'$ value moderately well, while 
B3LYP, mPW1PW91, B97, B97-1, HCTH,
and HCTH-120 all underestimate the central barriers.
The experimental central barrier for the F/Cl system 
is nearly 4 kcal/mol higher than the calculated values. Judging from the
performance of the various methods for the identity S$_N$2 reactions,
it is almost certain that the experimental F/Cl central 
barrier is in error and unambiguous new measurements are in order.
For the F/Cl and Cl/Br systems, the B3LYP, mPW1PW91, B97, B97-1, HCTH, 
and HCTH-120 central barriers are all underestimated, while these
exchange-correlation functionals find no barrier at all for the
F/Br case.
Like for the identity case, mPW1K/cc-pVTZ(+X) and BH\&HLYP/cc-pVTZ(+X)
central barriers agree well with the benchmark {\em ab initio} values, 
although the basis set sensitivity of particularly the Cl/Br results
argues against using small basis sets like 6-31+G*.

Several studies have reported experimental overall barrier heights 
($\Delta${\it H}$^\ddag$$_{ovr}$), but only for the Cl/Br system, and 
the experimental data range from -0.61 to -1.83 kcal/mol.
To our knowledge, no experimental data are available for the F/Cl and F/Br systems.
For the Cl/Br system,  the theoretical values span a range 
from -1.17 to -6.60 kcal/mol. Nevertheless, it is worth noting that
the W1$'$ value (-1.82 kcal/mol) for the 
Cl/Br system is in excellent agreement
with the experimental overall barrier height reported by Caldwell and 
coworkers\cite{caldwell84} (-1.83 kcal/mol, after accounting
for ZPVE and thermal corrections). Some caution should be exercised 
as the W1$'$ value does not include the core correlation contribution. 
Also note that G2 theory (-1.82 kcal/mol) reproduces the W1$'$ value 
very well. Except for mPW1K/cc-pVTZ(+X), mPWH\&HPW91/cc-pVTZ(+X),
and BH\&HLYP/cc-pVTZ(+X), all the DFT methods perform 
poorly, consistent with the preceding discussion.

\section{Conclusions}

A benchmark study using the W1$'$ and W2h methods has been carried out
for the potential surface of the gas-phase $S_N2$ reactions
Y$^-$ + CH$_3$X $\longrightarrow$ CH$_3$Y + X$^-$. A number of more 
approximate (and less expensive) methods --- both compound models (like
G2/G3 theory and CBS-QB3) and density functional methods ---
have been applied in an attempt to assess their performance for
barrier heights in $S_N2$ reactions. We arrive at the following conclusions.

\noindent {\bf (1)} Our best calculations are in excellent agreement with experiment for
the $\Delta H^\circ_f$ values of the methyl halides (where available)
and the electron affinities of the halogens. Where accurate experimental
data are available for the title reactions (e.g. for the Cl/Cl case), 
our best calculations agree with experiment to within overlapping uncertainties.
Our calculations however suggest that more reliable experimental data are
in order for most of the reactions considered.

\noindent {\bf (2)} The nonidentity S$_N$2 reactions and F/F identity reaction
possess transition state structures below the reactants energy
while Cl/Cl and Br/Br transition structures are above the reactants
energy. The complexation energies for identity S$_N$2 reactions are found
to increase in the order Br $<$ Cl $<$ F while the barrier heights follow
the order F $\lesssim$ Br $<$ Cl.
The complexation energies for the nonidentity S$_N$2 reactions indicate
that the $\Delta$$H$$_{comp}$ strongly depend on the nucleophile and leaving group.

\noindent {\bf (3)} The B3LYP,  and to a lesser extent, mPW1PW91 exchange-correlation 
functionals systematically underestimate barrier heights and, in the F/Br
case, are not even able to locate the correct stationary points on the 
potential surface. The latter problem is remedied by using the corresponding
`half-and-half' functionals BH\&HLYP and mPWH\&HPW91, which however appear
to somewhat overcorrect the barrier height. The B97 and B97-1 functionals
perform similarly to B3LYP for the problem under study.  The `pure DFT' HCTH and 
HCTH-120 functionals both underestimate central barrier heights; 
HCTH in addition underestimates complexation energies (and severely 
overestimates ion-molecule distances), which are
however well reproduced by HCTH-120.
Overall, the mPW1K functional appears to
put in the best performance of all DFT methods considered, especially
when using extended basis sets.

\noindent {\bf (4)} The performance of G2(+), G3(+), and CBS-QB3(+) methods for the 
energetics still appears to be superior to the DFT methods. (The `(+)'
stands for the addition of diffuse functions to the basis set used in 
obtaining the reference geometries; this is mandatory to get transition
states at all in the F/Br and Cl/Br cases.) The limitations for transition
states of the B3LYP exchange-correlation functional suggest its replacement
--- at least for kinetics applications ---
by mPW1K in thermochemistry methods that employ DFT reference geometries,
e.g. G3B3, CBS-QB3, and W1 theory. (In addition, larger basis sets than
6-31+G* should definitely be considered for the Br compounds.)

The present calculations illustrate the power of state-of-the-science
theoretical methods in providing both qualitative and quantitative
information regarding the reaction energetics. In the absence of
accurate experimental data, our high quality results should be
useful to future experimental and theoretical studies.

\acknowledgments
SP and GdO acknowledge Postdoctoral Fellowships from the Feinberg 
Graduate School (Weizmann Institute). JM is the incumbent of the Helen 
and Milton A. Kimmelman Career Development Chair. The authors would like to thank
Mark Iron for editorial assistance. This research was 
supported by the Minerva Foundation, Munich, Germany, and by the 
{\it Tashtiyot} Program of the Ministry of Science (Israel).
\\  \linebreak
\noindent{\bf Supporting Information Available:} Calculated
geometries of species involved in the S$_N$2 reactions
are available on the World Wide Web at the Uniform Resource Locator
(URL) http://theochem.weizmann.ac.il/web/papers/sn2.html.
This material is also available free of charge via the
Internet at http://pubs.acs.org.

\newpage
\begin{table}
\caption{\label{w1w2}Components of computed electron affinities of X and
total atomization energies (kcal mol$^{-1}$) of CH$_3$X (X = F, Cl and Br).}
\begin{tabular}{lrrrrrrr}
        & SCF   & CCSD  & (T)     & Core  & Spin-Orbit & Scalar Rel. & Final  \\
Species & limit & limit & limit   & corr. & splitting  &  effects    & Energy    \\
\hline
W1$'$    &      &      &      &      &      &      &                                                  \\
F$^-$   &   30.21   &   44.76   &   4.17   &   0.17   &   -0.39   &   -0.26   &   78.66             \\
Cl$^-$   &   58.35   &   24.16   &   2.28   &   0.02   &   -0.84   &   -0.34   &   83.63            \\
Br$^-$   &   58.37   &   21.76   &   1.98   &   0.31   &   -3.51   &   -0.90   &   78.02            \\
CH$_3$F   &   319.57   &   97.39   &   5.41   &   1.12   &   -0.47   &   -0.37   &   422.65         \\
CH$_3$Cl   &   303.73   &   86.14   &   5.27   &   1.19   &   -0.93   &   -0.42   &   394.98        \\
CH$_3$Br   &   292.99   &   85.48   &   5.25   &   1.45   &   -3.60   &   -0.79   &   380.78        \\
\hline
W2h    &      &      &      &      &      &      &                                                  \\
F$^-$   &  30.08   &  44.71   &  4.15   &  0.17   &  -0.39   &  -0.26   &  78.46                   \\
Cl$^-$   &  58.33   &  23.82   &  2.26   &  0.02   &  -0.84   &  -0.34   &  83.25                  \\
Br$^-$   &  58.31   &  21.49   &  1.91   &  0.31   &  -3.51   &  -0.90   &  77.62                  \\
CH$_3$F   &  319.82   &  96.89   &  5.34   &  1.14   &  -0.47   &  -0.37   &  422.34               \\
CH$_3$Cl   &  303.90   &  86.03   &  5.28   &  1.21   &  -0.93   &  -0.42   &  395.07              \\
CH$_3$Br   &  293.14   &  85.45   &  5.21   &  1.49   &  -3.60   &  -0.79   &  380.91              \\
\end{tabular}
\end{table}
\newpage

\begin{table}
\caption{\label{rel}Components of relative energies (kcal mol$^{-1}$) of ion-molecule complexes and transition
state structures with respect to reactants.}
\begin{tabular}{lrrrrrd}
        & SCF   & CCSD  & (T)     & Core  &  Scalar Rel. &  Final  \\
Species & limit & limit & limit   & corr. &  effects    & Energy    \\
\hline
W1$'$   &      &      &      &      &      &                                                  \\ 
F$^-$$\cdots$CH$_3$F             
&  -10.63   &   -2.54   &   -0.56   &   0.09   &   -0.02   &   -13.66    \\
$\lbrack$F$\cdots$CH$_3$$\cdots$F$\rbrack$$^-$   
&  8.23   &   -6.23   &   -2.65   &   0.36   &   -0.08   &   -0.37       \\
Cl$^-$$\cdots$CH$_3$Cl           
&   -8.08   &   -1.93   &   -0.62   &   0.08   &   0.01   &   -10.54     \\
$\lbrack$Cl$\cdots$CH$_3$$\cdots$Cl$\rbrack$$^-$ 
&  7.81   &   -2.62   &   -2.37   &   0.35   &   -0.10   &   3.07        \\
Br$^-$$\cdots$CH$_3$Br            
& -7.55   &   -1.92   &   -0.65   &           &   0.09   &   -10.03$^a$ \\
$\lbrack$Br$\cdots$CH$_3$$\cdots$Br$\rbrack$$^-$  
&   5.71   &   -2.43   &   -2.36   &   0.26   &   -0.16   &   1.02      \\
\hline
W2h   &      &      &      &      &      &                                                  \\
F$^-$$\cdots$CH$_3$F           
& -11.16   &   -2.17   &   -0.45   &   0.07   &   -0.02   &   -13.72       \\
$\lbrack$F$\cdots$CH$_3$$\cdots$F$\rbrack$$^-$  
&  8.58   &   -6.61   &   -2.58   &   0.35   &   -0.08   &   -0.34        \\
Cl$^-$$\cdots$CH$_3$Cl   
&   -8.01   &   -2.39   &   -0.65   &   0.10   &   0.01   &   -10.94             \\
$\lbrack$Cl$\cdots$CH$_3$$\cdots$Cl$\rbrack$$^-$   
& 8.55   &   -3.63   &   -2.46   &   0.33   &   -0.11   &   2.67        \\ 
$\lbrack$Br$\cdots$CH$_3$$\cdots$Br$\rbrack$$^-$   
&  6.57   &   -3.41   &   -2.38   &   0.21   &   -0.21   &   0.77       \\
\hline
W1$'$   &      &      &      &      &      &                                                   \\
F$^-$$\cdots$CH$_3$Cl   
&  -13.67   &   -1.16   &   -0.76   &   0.18   &   -0.02   &   -15.43             \\
$\lbrack$F$\cdots$CH$_3$$\cdots$Cl$\rbrack$$^-$  
&  -10.39   &   -0.84   &   -1.60   &   0.36   &   -0.07   &   -12.54    \\
Cl$^-$$\cdots$CH$_3$F   
&   -7.08   &   -2.03   &   -0.42   &   0.03   &   0.00   &   -9.51               \\
F$^-$$\cdots$CH$_3$Br   
&  -15.08   &   -1.24   &   -0.72   &     &   0.03   &   -17.01$^a$                \\
$\lbrack$F$\cdots$CH$_3$$\cdots$Br$\rbrack$$^-$ 
&  -13.85   &   -1.02   &   -1.43   &    &   -0.07   &   -16.37$^a$      \\
Br$^-$$\cdots$CH$_3$F   
&  -6.63   &   -1.69   &   -0.24   &    &   0.05   &   -8.51$^a$                  \\
Cl$^-$$\cdots$CH$_3$Br   
&  -8.89   &   -2.02   &   -0.59   &      &   -0.41   &   -11.91$^a$       \\
$\lbrack$Cl$\cdots$CH$_3$$\cdots$Br$\rbrack$$^-$   
&  2.24   &   -2.69   &   -2.28   &      &   -0.57   &   -3.30$^a$   \\
Br$^-$$\cdots$CH$_3$Cl   
&  -7.38   &   -2.03   &   -0.45   &      &   -0.46   &   -10.32$^a$                 \\
\end{tabular}
$^a$ W1$'$ $-$ Core Correlation 
\end{table}
\newpage

\begin{table}
\caption{\label{hof}Calculated and experimental heats of formation (kcal mol$^{-1}$) of CH$_3$X and
electron affinities (eV) of X (X = F, Cl and Br)}
\begin{tabular}{lrrrlrrr}
          &  \multicolumn{3}{c}{Heats of Formation} & & \multicolumn{3}{c}{Electron Affinity}   \\
          & \multispan3{\hrulefill} & & \multispan3{\hrulefill} \\
Species   &  W1$'$   &   W2h   & Experiment &  Species   &  W1'   &   W2h   & Experiment   \\
\hline
CH$_3$F   &  -57.06 & -56.75 &  -56(7)\cite{janaf}  &  F    &  3.411   & 3.402  &  3.401 190(4)\cite{crc}   \\
CH$_3$Cl  &  -20.14 & -20.23 &  -20.00(50)\cite{janaf}  &  Cl   &  3.627   & 3.610  &  3.612 69(6)\cite{crc}    \\
CH$_3$Br  &  -8.50  & -8.63  &  -8.20(19)\cite{webbook}   &  Br   &  3.383   & 3.366  &  3.363 583(44)\cite{webbook}   \\
\end{tabular}
\end{table}

\newpage

\begin{table}  
\caption{\label{id}Comparison of complexation energies ($\Delta${\it H}$_{comp}$) 
of the ion-molecule complexes, overall barrier heights relative to reactants
($\Delta${\it H}$^\ddag$$_{ovr}$), and central barriers ($\Delta${\it H}$^\ddag$$_{cent}$) 
of identity S$_N$2 reactions, X$^-$ + CH$_3$X $\longrightarrow$ XCH$_3$ + X$^-$, 
calculated at various levels of theory. All values in kcal/mol.}
{
\squeezetable
\begin{tabular}{llddd}
  X    &    Method     &  $\Delta${\it H}$_{comp}$  & $\Delta${\it H}$^\ddag$$_{ovr}$ &
                                     $\Delta${\it H}$^\ddag$$_{cent}$   \\
\hline
F &  W1$'$  &  13.66  &  -0.37  &  13.29            \\   
  &  W2h  &  13.72  &  -0.34  &  13.38            \\
  &  B3LYP/cc-pVTZ(+X)  &  12.72  &  -2.58  &  10.15          \\
  &  BH\&HLYP/cc-pVTZ(+X)  &  13.22  &   1.31  &  14.53          \\
  &  mPW1PW91/cc-pVTZ(+X)  &  12.49  &  -0.95  &  11.55       \\
  &  mPWH\&HPW91/cc-pVTZ(+X)  &  12.77  &   2.60  &  15.38       \\
  &  B97/cc-pVDZ(+X)  &  12.48 &  -2.47  &  10.01           \\
  &  B97-1/cc-pVDZ(+X) &  13.21 &  -3.29  &   9.92            \\
  &  HCTH/cc-pVDZ(+X)  &  9.87   &  -0.60   & 9.27         \\
&  HCTH-120/cc-pVDZ(+X)&  12.39 & -4.20 & 8.18 \\
  &  mPW1K/6-31+G*  &  13.55  &  -0.30  &  13.26          \\
  &  mPW1K/cc-pVDZ(+X) &  12.63  &  0.36  &  13.00       \\
  &  mPW1K/cc-pVTZ(+X)  &  12.66  &  1.69  &  14.36        \\
  &  G1  &  13.01  &  -1.37  &  11.64              \\
  &  G2  &  13.34  &  -1.15  &  12.19              \\
  &  G2MP2  &  13.41  &  -0.63  &  12.78           \\
  &  G3  &  14.23  &  -1.97  &  12.26              \\
  &  CBS-QB3  &  13.46  &  -0.85  &  12.61             \\
  &  G3(+)  &  14.59  &  -2.68  &  11.90            \\
  &  CBS-QB3(+)  &  14.15  &  -0.52  &  13.63           \\
  &  G2(+)$^a$  &  13.81  &  -1.86  &  11.95            \\
	&  CCSD(T)/$spdf$$^c$ & 13.73 & -0.92 & 12.81 \\
\hline
Cl&  W1$'$  &  10.54  &  3.07  &  13.61               \\
  &  W2h  &  10.94  &  2.67  &  13.61               \\  
  &  B3LYP/cc-pVTZ(+X)  &  9.50  &  -0.48  &  9.02              \\ 
  &  BH\&HLYP/cc-pVTZ(+X)  &  9.67  &   3.17  &  12.84             \\ 
  &  mPW1PW91/cc-pVTZ(+X)  &  9.59  &  1.23  &  10.82           \\ 
  &  mPWH\&HPW91/cc-pVTZ(+X)  &  9.69  &  4.50  &  14.19           \\ 
  &  B97/cc-pVDZ(+X)  &  10.10 &  -0.66  &   9.44           \\
  &  B97-1/cc-pVDZ(+X) &  10.74 & -1.46  &  9.28           \\
  &  HCTH/cc-pVDZ(+X)  &  7.91   &  1.45   & 9.36         \\
& HCTH-120/cc-pVDZ(+X) &  9.96 & -1.93 & 8.03 \\
  &  mPW1K/6-31+G*  &  9.75  &  3.20  &  12.95              \\ 
  &  mPW1K/cc-pVDZ(+X)  &  9.65  &  3.63  &  13.28          \\ 
  &  mPW1K/cc-pVTZ(+X)  &  9.64  &  3.66  &  13.30          \\ 
  &  G1  &  10.52  &  1.79  &  12.31                 \\
  &  G2  &  10.77  &  3.06  &  13.83                 \\
  &  G2MP2  &  10.89  &  2.74  &  13.63              \\
  &  G3  &  11.15  &  1.79  &  12.95                 \\
  &  CBS-QB3  &  10.65  &  2.47  &  13.12                \\
  &  G3(+)  &  11.04  &  1.80  &  12.83              \\
  &  CBS-QB3(+)  &  10.69  &  2.40  &  13.09              \\
  &  G2(+)$^a$  &  10.71  &  3.01  &  13.72               \\
  &  CCSD(T)/$spdfg$$^b$  &    &  2.75  &                         \\
  &  Experiment  &  10.53(40)$^d$ &  2.90$^e$ &  13.66(2.01)$^f$       \\
\hline
  &    &    &    &                                   \\ 
Br&  W1$'$  $-$ core  &  10.03  &  1.02$^i$  &  10.79        \\
  &  W2h   &         &  0.77  &          \\
  &  B3LYP/cc-pVTZ(+X) &  9.06  &  -2.42  &  6.64        \\
  &  BH\&HLYP/cc-pVTZ(+X) &  9.04  &   1.25  &  10.29       \\
  &  mPW1PW91/cc-pVTZ(+X)  &  9.21  &  -1.03  &  8.18    \\
  &  mPWH\&HPW91/cc-pVTZ(+X)  &  9.19  &  2.22 &  11.40    \\
  &  B97/cc-pVDZ(+X)  &  9.62 &  -2.29  &   7.33          \\
  &  B97-1/cc-pVDZ(+X) &  10.24 &  -3.02  &   7.22           \\
  &  HCTH/cc-pVDZ(+X)  &  7.56   &  -0.70   & 6.86         \\
  &  HCTH-120/cc-pVDZ(+X) & 9.73 & -4.06 & 5.68 \\
  &  mPW1K/6-31+G*  &  12.78  &  -1.95  &  10.83         \\
  &  mPW1K/cc-pVDZ(+X) &  9.34  &  0.68  &  10.02        \\
  &  mPW1K/cc-pVTZ(+X) &  9.16  &  1.38  &  10.54        \\
  &  G1  &  9.68  &  1.11  &  10.78                      \\
  &  G2  &  9.85  &  1.52  &  11.38                      \\
  &  G2MP2  &  9.83  &  1.83  &  11.66                   \\
  &  G2(+)$^a$ $-$ ECP &  10.17  &  1.48  &  11.65       \\
  & Experiment   & 11.34(40)$^d$ & 1.73$^g$ & 11.68$^h$        \\
\end{tabular}
}
$^a$ G2(+) values are from Ref.\cite{radomid95};
$^b$ CCSD(T)/$spdfg$ values are from Ref.\cite{botsch98};
$^c$ Ref.\cite{Wla94}\\
Experimental values:  $^d$From Ref.\cite{li96},          
$^e$ From Ref.\cite{wladkowski93},  
$^f$ From Ref.\cite{bierbaum88},   
$^g$ From Ref.\cite{wilbur93},    
$^h$ From Ref.\cite{pellerite83}     \\
$^i$ Core contribution included\\
\end{table}
\newpage

\begin{table}
\caption{\label{complex} Comparison of complexation energies ($\Delta${\it H}$_{comp}$, kcal/mol) of 
the ion-molecule complexes for the nonidentity S$_N$2 reactions, calculated at various levels of theory.}
\bigskip \setlength{\columnsep}{0.0pt}
\squeezetable
\begin{tabular}{ldddddd}
Method & F$^-$$\cdots$CH$_3$Cl  &  Cl$^-$$\cdots$CH$_3$F  &  F$^-$$\cdots$CH$_3$Br  
&  Br$^-$$\cdots$CH$_3$F  &  Cl$^-$$\cdots$CH$_3$Br  &  Br$^-$$\cdots$CH$_3$Cl \\
\hline
W1$'$ $-$ core  &  15.43$^a$  &  9.51$^a$  &  17.01  &  8.51  &  11.91  &  10.32       \\           
B3LYP/cc-pVTZ(+X)  &  15.37  &  8.09  &    & 7.18   &  10.24  &  8.42                         \\ 
BH\&HLYP/cc-pVTZ(+X) & 15.39 & 8.42  & 16.65 & 7.45  &10.23 &  8.56   \\
mPW1PW91/cc-pVTZ(+X)  &  15.06  &  8.13  &    & 7.27  &  10.32  &  8.54                      \\ 
mPWH\&HPW91/cc-pVTZ(+X)  & 15.03  & 8.32  & 16.28  &  7.42  &  10.30   &  8.65  \\
B97/cc-pVDZ(+X)  &  15.43  & 8.49  &      &   7.69  &  10.66  & 9.13    \\ 
B97-1/cc-pVDZ(+X) & 16.26  &  9.05  &       & 8.21  & 11.32  & 9.74    \\
HCTH/cc-pVDZ(+X)  &  12.66  &  6.56  &      &   5.82  & 8.56  &  7.00   \\
HCTH-120/cc-pVDZ(+X) & 15.42 & 8.48 &     & 7.69 & 10.78 & 9.01\\
mPW1K/6-31+G* &  15.32  &  8.79  &  17.63  &  9.12  &  12.97  &  10.56               \\ 
mPW1K/cc-pVDZ(+X)  &  14.52  &  8.46  &  16.25  &  7.62  &  10.41  &  8.68             \\
mPW1K/cc-pVTZ(+X) &  14.97  &  8.25  &  16.30  &  7.37  &  10.27  &  8.60             \\
G1  &    &  9.58  &    &  8.35  &  11.17  &  9.10                                  \\
G2  &    &  9.68  &    &  8.42  &  11.35  &  9.36                                  \\
G2MP2  &    &  9.71  &    &  8.40  &  11.37  &  9.45                               \\
G3  &    &  10.03  &    &    &    &                                            \\
CBS-QB3  &    &  9.33  &    &    &    &                                             \\
G3(+)  &  16.34  &  9.97  &    &    &    &                                      \\
CBS-QB3(+)  &  15.85  &  9.51  &    &    &    &                                     \\
G2(+)$^b$  &  15.62  &  9.64  &  16.74  &  8.56  &  11.47  &  9.64                  \\
CCSD(T)/large$^c$  &  16.07  &  9.75  &    &    & 11.31   & 9.71                                 \\
Experiment.    &    &  11.41(2.01)$^d$   &    &   &  12.54(40)$^e$   & 11.01 (40)$^e$      \\
\end{tabular}
$^a$ Core contribution included   \\
$^b$ G2(+) values are from Ref.\cite{radomnonid96},  \\
$^c$ F/Cl: CCSD(T)/$spdfg$ values from Ref.\cite{botsch97}; Cl/Br: CCSD(T)/$spdfgh$ values
from Ref.\cite{Sch2000}.\\
Experimental values:  $^d$From Ref.\cite{larson84},
$^e$ From Ref.\cite{li96}
\end{table}
\newpage

\begin{table}
\caption{\label{nonid}
Comparison of overall reaction enthalpies ($\Delta${\it H}$_{ovr}$), central 
enthalpy differences between reactant and product ion-molecule complexes 
($\Delta${\it H}$_{cent}$), overall barrier heights 
($\Delta${\it H}$^\ddag$$_{over}$) and central barrier 
heights ($\Delta${\it H}$^\ddag$$_{cent}$) for exothermic 
Y$^-$ + CH$_3$X $\longrightarrow$ YCH$_3$ + X$^-$
reactions, calculated at various levels of theory. All values in kcal/mol.}
\squeezetable
\begin{tabular}{lldddd}
  Y,X    &    Method     & $\Delta${\it H}$_{ovr}$  
                           &  $\Delta${\it H}$_{cent}$
                              &  $\Delta${\it H}$^\ddag$$_{over}$
                                &  $\Delta${\it H}$^\ddag$$_{cent}$   \\
\hline
F/Cl  &  W1$'$   &  -32.65  &  -26.73  &  -12.54  &  2.89                \\  
  &  B3LYP/cc-pVTZ(+X)  &  -32.77  &  -25.49  &  -14.69  &  0.67                  \\ 
  &  BH\&HLYP/cc-pVTZ(+X) & -37.02  &  -30.05  &  -12.86  &  2.53   \\
  &  mPW1PW91/cc-pVTZ(+X)  &  -33.08  &  -26.15  &  -13.43  &  1.63               \\ 
  & mPWH\&HPW91/cc-pVTZ(+X)  & -36.45  &  -29.73  &  -11.51  &  3.52      \\
  &  B97/cc-pVDZ(+X)  &  -32.90 &  -25.95 &  -14.70  & 0.73    \\
  &  B97-1/cc-pVDZ(+X) & -33.13 &  -25.92  &   -15.60    &  0.66    \\
  &  HCTH/cc-pVDZ(+X)  &  -30.77   & -24.67  & -11.95 & 0.71        \\
  &HCTH-120/cc-pVDZ(+X)&  -30.58 & -23.64 & -15.14 & 0.27 \\
  &  mPW1K/6-31+G*  &  -36.59  &  -30.07  &  -13.02  &  2.30                  \\ 
  &  mPW1K/cc-pVDZ(+X)  &  -34.74  &  -28.68  &  -11.95  &  2.57               \\
  &  mPW1K/cc-pVTZ(+X) &  -35.50  &  -28.78  &  -11.97  &  3.01               \\
  &  G1  &  -30.62  &    &    &                                        \\
  &  G2  &  -31.59  &    &    &                                        \\
  &  G2MP2  &  -32.25  &    &    &                                     \\
  &  G3  &  -33.00  &    &    &                                        \\
  &  CBS-QB3  &  -35.21  &    &    &                                        \\
  &  G3(+)  &  -32.86  &  -26.50  &  -14.04  &  2.30                    \\
  &  CBS-QB3(+)  &  -35.15  &  -28.81  &  -13.77  &  2.07                   \\
  &  G2(+)$^a$  &  -31.44  &  -25.46  &  -12.63  &  2.98                    \\
  &  CCSD(T)/$spdfg^b$  &  -32.34  &  -26.36  &  -11.84  &  3.89                   \\
  &  Experiment  &  -33.34(72)$^c$  &    &    &  7.52(1.20)$^d$                                    \\ 
  &    &    &    &    &                                                    \\           
\hline
F/Br  &  W1$'$ $-$ core  &  -41.43  &  -32.93  &  -16.37  &  0.64                     \\ 
  &  B3LYP/cc-pVTZ(+X) &  -40.78  &    &    &                                          \\
  &  BH\&HLYP/cc-pVTZ(+X) &  -46.13  &  -36.93  &  -16.33  &  0.32    \\
  &  mPW1PW91/cc-pVTZ(+X)  &  -41.65  &    &    &                                       \\
  &  mPWH\&HPW91/cc-pVTZ(+X)  & -45.90  & -37.05  &  -15.42 &  0.85      \\
  &  B97/cc-pVDZ(+X)  &  -40.41 &       &       &          \\
  &  B97-1/cc-pVDZ(+X) & -40.68 &       &        &               \\
  &  HCTH/cc-pVDZ(+X)  &  -38.18  &      &       &    \\
  &HCTH-120/cc-pVDZ(+X)&  -37.96  &     &     &   \\
  &  mPW1K/6-31+G* &  -42.35  &  -33.84  &  -16.92  &  0.71                        \\
  &  mPW1K/cc-pVDZ(+X)  &  -44.76  &  -36.13  &  -16.08  &  0.18                    \\
  &  mPW1K/cc-pVTZ(+X) &  -44.70  &  -35.77  &  -15.76  &  0.54                    \\
  &  G1  &  -40.87  &    &    &                                             \\
  &  G2  &  -40.02  &    &    &                                              \\
  &  G2MP2  &  -40.35  &    &    &                                           \\
  &  G2(+)$^a$  &  -39.47  &  -31.29  &  -15.90  &  0.84                         \\
  &  Experiment  &  -40.20(96)$^c$  &    &    &                                            \\
\hline
Cl/Br  &  W1$'$ $-$ core  &  -8.56  &  -6.97  &  -1.82  &  8.61                        \\   
  &  B3LYP/cc-pVTZ(+X)  &  -8.01  &  -6.19  &  -5.25  &  4.99                            \\  
  &  BH\&HLYP/cc-pVTZ(+X) &  -9.11  &  -7.43  &  -2.15  &  8.08               \\
  &  mPW1PW91/cc-pVTZ(+X) &  -8.57  &  -6.79  &  -3.99  &  6.33                         \\  
  &  mPWH\&HPW91/cc-pVTZ(+X)  & -9.45  &  -7.79  &  -1.17  &  9.14         \\
  &  B97/cc-pVDZ(+X)  &  -7.52  &  -5.99  &  -5.06  &  5.60     \\
  &  B97-1/cc-pVDZ(+X) & -7.54 & -5.97  &  -5.83    &  5.48         \\
  &  HCTH/cc-pVDZ(+X)  &  -7.41   &  -5.85   & -3.18  &  5.38         \\
  &HCTH-120/cc-pVDZ(+X)&  -7.38 & -5.62 & -6.52 & 4.26 \\
  &  mPW1K/6-31+G*  &  -5.75  &  -3.34  &  -3.12  &  9.85                            \\  
  &  mPW1K/cc-pVDZ(+X)  &  -10.02  &  -8.30  &  -6.60  &  3.81                       \\  
  &  mPW1K/cc-pVTZ(+X)  &  -9.20  &  -7.53  &  -1.88  &  8.38                        \\  
  &  G1  &  -10.25  &  -8.18  &  -3.45  &  7.72                              \\  
  &  G2  &  -8.43  &  -6.44  &  -1.82  &  9.53                                \\ 
  &  G2MP2  &  -8.11  &  -6.18  &  -1.67  &  9.70                             \\ 
  &  G2(+)$^a$  &  -8.04  &  -6.21  &  -1.71  &  9.76                             \\ 
  & CCSD(T)/$spdfgh$$^{b}$ & -8.53 & -6.93 & -2.33 & 8.98 \\
  &  Experiment  &  -6.86(72)$^c$  &    &  -1.83(5)$^e$  &                                         \\ 
  &    &    &    &  -1.69(33)$^f$  &                                                   \\ 
   &   &    &    &  -1.52$^g$  &                                                   \\ 
  &   &    &    &  -1.11$^h$  &                                                   \\
 &   &    &    &  -0.61$^i$  &                                                    \\ 
\end{tabular}
$^a$ G2(+) values are from Ref.\cite{radomnonid96} \\
$^b$ Ref.\cite{botsch97} (F/Cl) and Ref.\cite{Sch2000} (Cl/Br)\\
$^{b'}$ Table 2 of Ref.\cite{Sch2000}\\
Experimental values:  $^c$From Ref.\cite{lias88},
$^d$ From Ref.\cite{pellerite83},
$^e$ From Ref.\cite{caldwell84},
$^f$ From Ref.\cite{graul94},
$^g$ From Ref.\cite{knighton93},
$^h$ From Ref.\cite{wladkowski93},
$^i$ From Ref.\cite{hu95}

\end{table}
\newpage

\begin{figure}[H]
\caption{\label{sn2fig}%
Schematic representation of potential energy surface for the (a) identity and 
(b) nonidentity S$_N$2 reactions
}
\end{figure}

\newpage
\thispagestyle{empty}
\vspace*{2in}
\centerline{\includegraphics[scale=1.0]{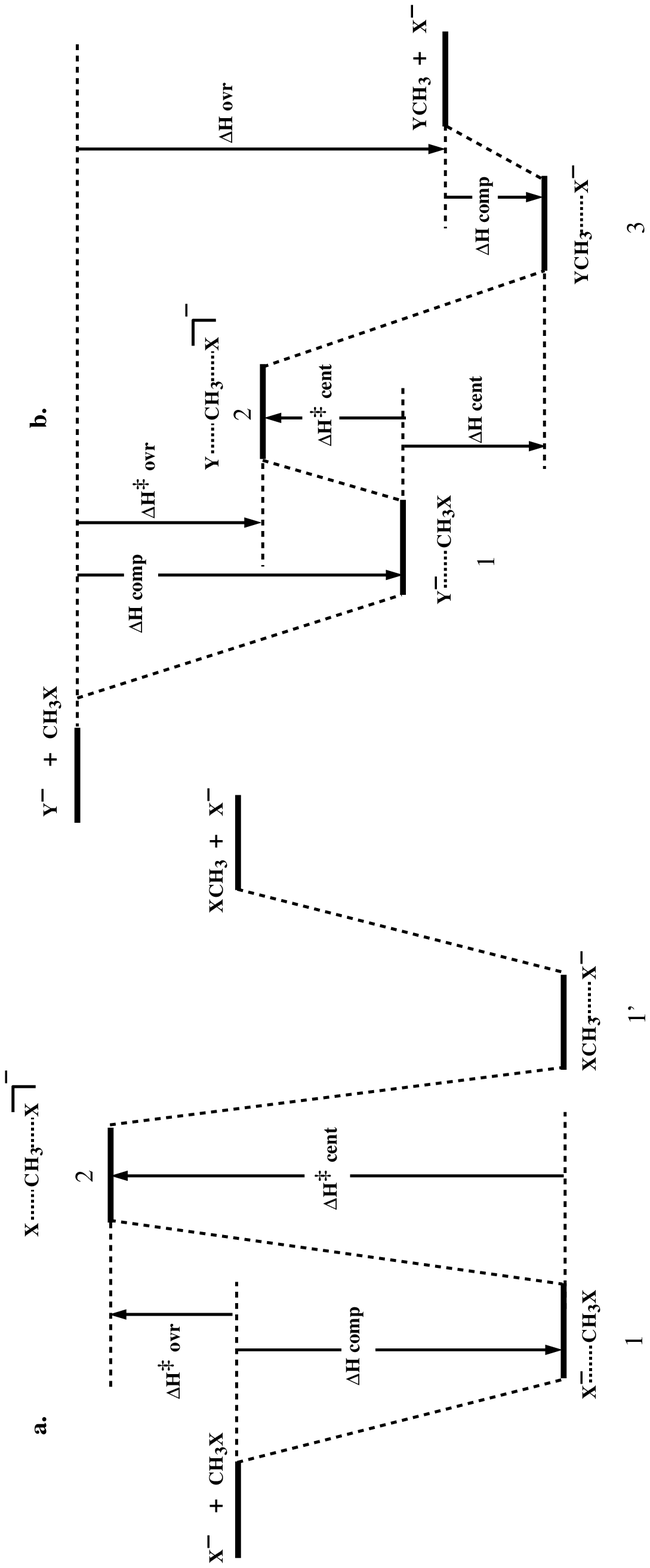}}

\end{document}


\draft
\title{Benchmark {\em ab initio} energy profiles for the gas-phase
S$_N$2 reactions Y$^-$ + CH$_3$X $\rightarrow$ CH$_3$Y + X$^-$
(X,Y = F,Cl,Br). Validation of hybrid DFT methods\\ \texttt{\large Supplementary Data}}

\author{Srinivasan Parthiban, Gl\^{e}nisson de Oliveira\thanks{Present address:
Chemistry Department, Pensacola Christian College, 250 Brent Lane, Pensacola, FL 32503}, and
Jan M.L. Martin\thanks{Author to whom correspondence should be addressed. Email: {\tt comartin@wicc.weizmann.ac.il}}}
\address{Department of Organic Chemistry,
Kimmelman Building, Room 262,
Weizmann Institute of Science,
IL-76100 Re\d{h}ovot, Israel.
}
\date{{\em J. Phys. Chem. A} manuscript JP0031000; revised October 31, 2000}
\maketitle

\newpage

\begin{table}
\caption{\label{ch3x}Calculated and experimental geometries (\AA\ , degree) of 
CH$_3$X (X=F, Cl, and Br).}
\squeezetable
\begin{tabular}{llrrr}
Species & Method/Basis Set & r(C$-$X)   & r(C$-$H) &  $\angle$XCH \\
\hline
CH$_3$F   &  CCSD(T)/cc-pVQZ+1      & 1.382   & 1.089   & 108.9   \\
          &  B3LYP/cc-pVTZ(+X)      & 1.392   & 1.090   & 108.7    \\
          &  BH\&HLYP/cc-pVTZ(+X)   & 1.372   & 1.082   & 108.8    \\
          &  mPW1PW91/cc-pVTZ(+X)   & 1.380   & 1.090   & 108.9    \\
          &  mPWH\&HPW91/cc-pVTZ(+X) & 1.362 & 1.083   & 109.1    \\
          &  B97/cc-pVDZ(+X)        & 1.399  &  1.102 &  108.7    \\
          &  B97-1/cc-pVDZ(+X)      & 1.398  &  1.102  &  108.7   \\
          &  HCTH/cc-pVDZ(+X)       & 1.400  &  1.103  &  108.8   \\
          &  HCTH-120/cc-pVDZ(+X)   & 1.402  &  1.104  &  108.7   \\
          &  mPW1K/6-31+G*          & 1.374   & 1.087   & 108.7    \\
          &  mPW1K/cc-pVDZ(+X)      & 1.378   & 1.093   & 108.8    \\
          &  mPW1K/cc-pVTZ(+X)      & 1.367   & 1.085   & 109.0    \\
          &  MP2/6-31G*             & 1.390   & 1.092   & 109.1    \\
          &  MP2/6-31+G*$^a$        & 1.407   & 1.090   & 108.0    \\
          &  B3LYP/6-311G(2d,d,p)   & 1.388   & 1.093   & 109.2    \\
          &  B3LYP/6-311+G(2d,d,p)  & 1.396   & 1.091   & 108.6    \\
          &  Expt.$^b$                  & 1.383   & 1.086   & 108.8    \\
\hline
CH$_3$Cl  &  CCSD(T)/cc-pVQZ+1     &  1.783 & 1.085  &  108.4    \\ 
          &  B3LYP/cc-pVTZ(+X)     & 1.796  & 1.085   &  108.3   \\
          &  BH\&HLYP/cc-pVTZ(+X)  & 1.779  & 1.077  &  108.4    \\
          &  mPW1PW91/cc-pVTZ(+X)   & 1.776  & 1.085   &  108.5   \\
          &  mPWH\&HPW91/cc-pVTZ(+X) & 1.761 & 1.079  &  108.6    \\
          &  B97/cc-pVDZ(+X)        & 1.808  &  1.097 &  108.1      \\
          &  B97-1/cc-pVDZ(+X)      & 1.808  &  1.097  & 108.0    \\
          &  HCTH/cc-pVDZ(+X)       & 1.793  &  1.098  &  108.4   \\
          &  HCTH-120/cc-pVDZ(+X)   &  1.796  & 1.098  &  108.4  \\
          &  mPW1K/6-31+G*          & 1.772   & 1.084   &  108.8   \\
          &  mPW1K/cc-pVDZ(+X)      & 1.768   & 1.090   & 108.5    \\
          &  mPW1K/cc-pVTZ(+X)      & 1.765   & 1.081   & 108.6    \\
          &  MP2/6-31G*             & 1.777   & 1.088   & 108.9    \\
          &  MP2/6-31+G*$^a$    & 1.780   & 1.089   & 108.9    \\
          &  B3LYP/6-311G(2d,d,p)   & 1.803   & 1.087   & 108.2    \\
          &  B3LYP/6-311+G(2d,d,p)  & 1.803   & 1.087   & 108.2    \\
          &  Expt.$^c$                  & 1.776   & 1.085   & 108.6    \\
\hline
CH$_3$Br  &  CCSD(T)/cc-pVQZ+1     & 1.944    & 1.084   &  107.8  \\
          &  B3LYP/cc-pVTZ(+X)      & 1.957   & 1.083   &  107.6   \\
          &  BH\&HLYP/cc-pVTZ(+X)  & 1.938    &  1.076   & 107.8    \\
          &  mPW1PW91/cc-pVTZ(+X)   & 1.936   & 1.084   & 107.9    \\
          &  mPWH\&HPW91/cc-pVTZ(+X) & 1.917 & 1.078   & 108.0    \\
          &  B97/cc-pVDZ(+X)        & 1.959   &  1.097  &  107.6   \\
          &  B97-1/cc-pVDZ(+X)      &  1.959  & 1.097  &  107.6   \\
          &  HCTH/cc-pVDZ(+X)       & 1.952  &  1.097  &  107.9   \\
          &  HCTH-120/cc-pVDZ(+X)   & 1.956  &  1.097  &  107.8   \\
          &  mPW1K/6-31+G*          & 1.925   & 1.083   & 108.2    \\
          &  mPW1K/cc-pVDZ(+X)      & 1.927   & 1.089   & 107.9    \\
          &  mPW1K/cc-pVTZ(+X)      & 1.922   & 1.079   & 108.0    \\
          &  MP2/6-31G*             & 1.947   & 1.086   & 107.9    \\
          &  MP2/6-31+G*$^a$    & 1.954   & 1.088   & 108.0    \\
          &  Expt.$^d$                  & 1.934   & 1.082   & 107.7    \\
\end{tabular}
$^a$ From Glukhovtsev, M.N.; Pross, A.;  Radom, L.;
{\em J. Am. Chem. Soc.} {\bf 1995}, {\it 117}, 2024. \\
Experimental values: \\
$^b$ From Egawa, T.; Yamamoto, S.; Nakata, M.; Kuchitsu, K.;
{\em J. Mol. Struct.}{\bf 1987}, {\it 156}, 213.  \\
$^c$ From Jensen, T.; Brodersen, S.; Guelachvili, G.; {\em J. Mol.Spectrosc.}
{\bf 1981}, {\it 88}, 378.   \\
$^d$ From Graner, G.; {\em J. Mol.Spectrosc.} {\bf 1981}, {\it 90}, 394.

\end{table}

\newpage

\begin{table}
\caption{\label{x-ch3x}Geometries (\AA\ , degree) of ion-molecule complexes X$^-$$\cdots$CH$_3$X
(X=F, Cl, and Br) of the SN$_2$ identity reactions.}
\squeezetable
\begin{tabular}{llrrrr}
Species & Method/Basis Set & r(X$\cdots$C)   & r(C$-$X) & r(C$-$H)  & $\angle$HCX \\
\hline
X=F       & CCSD(T)/cc-pVQZ+1       & 2.494  & 1.432  & 1.082  & 109.1  \\
          &  B3LYP/cc-pVTZ(+X)      & 2.588  & 1.447  & 1.082  & 108.4  \\
          &  BH\&HLYP/cc-pVTZ(+X)  &  2.581  & 1.418  & 1.075  & 108.8  \\
          &  mPW1PW91/cc-pVTZ(+X)   & 2.575  & 1.428  & 1.083  & 108.9  \\
          &  mPWH\&HPW91/cc-pVTZ(+X) & 2.567 & 1.404 & 1.077  & 109.2  \\
          &  B97/cc-pVDZ(+X)        &  2.643  & 1.453 &  1.095 & 108.3  \\
          &  B97-1/cc-pVDZ(+X)       &  2.613  & 1.453 &  1.095 &  108.3  \\
          &  HCTH/cc-pVDZ(+X)        & 2.794  & 1.453  &  1.096 &  108.4  \\
          &  HCTH-120/cc-pVDZ(+X)     & 2.668  & 1.462  &  1.096 &  108.2  \\
          &  mPW1K/6-31+G*          & 2.572  & 1.421  & 1.080  & 108.5  \\
          &  mPW1K/cc-pVDZ(+X)      & 2.579  & 1.425  & 1.087  & 108.8   \\
          &  mPW1K/cc-pVTZ(+X)      & 2.571  & 1.411  & 1.079  & 109.1 \\
          &  MP2/6-31G*             & 2.426  & 1.439  & 1.083  & 109.5  \\
          &  MP2/6-31+G*$^a$        & 2.628  & 1.456  & 1.084  & 107.7   \\
          &  B3LYP/6-311G(2d,d,p)   & 2.420  & 1.453  & 1.084  & 110.0   \\
          &  B3LYP/6-311+G(2d,d,p)  & 2.565  & 1.455  & 1.084  & 108.2   \\
\hline                                      
X=Cl      &  CCSD(T)/cc-pVQZ+1    & 3.123   &  1.860  & 1.080  & 107.8  \\
          &  B3LYP/cc-pVTZ(+X)      & 3.191 &  1.846  & 1.080  & 107.9  \\
          &  BH\&HLYP/cc-pVTZ(+X)  &  3.203 &  1.818  & 1.073  & 108.2  \\
          &  mPW1PW91/cc-pVTZ(+X)   & 3.163 &  1.817  & 1.081  & 108.5   \\
          &  mPWH\&HPW91/cc-pVTZ(+X) & 3.166 & 1.796 & 1.075 & 108.7   \\
          &  B97/cc-pVDZ(+X)        &  3.198  &  1.855 &  1.094 & 107.7  \\
          &  B97-1/cc-pVDZ(+X)      &  3.168 & 1.857 &  1.093  &  107.6  \\
          &  HCTH/cc-pVDZ(+X)       &  3.375  & 1.838 & 1.094 &  108.1   \\
          &  HCTH-120/cc-pVDZ(+X)   &  3.241 &  1.848 &  1.095 & 107.9  \\
          &  mPW1K/6-31+G*          & 3.181 &  1.809  & 1.080  & 108.7   \\
          &  mPW1K/cc-pVDZ(+X)      & 3.182 &  1.804  & 1.086  & 108.6   \\
          &  mPW1K/cc-pVTZ(+X)      & 3.167 &  1.801  & 1.077  & 108.7   \\
          &  MP2/6-31G*             & 3.158 &  1.812  & 1.084  & 109.0   \\
          &  MP2/6-31+G*$^a$    & 3.270 &  1.810  & 1.085  & 108.8  \\
          &  B3LYP/6-311G(2d,d,p)   & 3.146 &  1.861  & 1.082  & 107.6  \\
          &  B3LYP/6-311+G(2d,d,p)  & 3.187 &  1.854  & 1.083  & 107.8   \\
\hline                                      
X=Br      &  CCSD(T)/cc-pVQZ+1      & 3.277  & 2.023  & 1.079  & 106.9  \\
          &  B3LYP/cc-pVTZ(+X)      & 3.321  & 2.016  & 1.079  & 106.9  \\
          &  BH\&HLYP/cc-pVTZ(+X)  & 3.361  & 1.980 &  1.072  &  107.4   \\
          &  mPW1PW91/cc-pVTZ(+X)   & 3.290  & 1.982  & 1.080  & 107.5   \\
          &  mPWH\&HPW91/cc-pVTZ(+X) & 3.313 & 1.953  &  1.074 & 107.9 \\
          &  B97/cc-pVDZ(+X)        &  3.320  &  2.013 &  1.093 & 107.0  \\
          &  B97-1/cc-pVDZ(+X)      &  3.281  &  2.014 &  1.093 &  106.9  \\
          &  HCTH/cc-pVDZ(+X)       &  3.471  &  2.007  &  1.094 &  107.2 \\
          &  HCTH-120/cc-pVDZ(+X)   &  3.332 &  2.022 &  1.094 &  106.9  \\
          &  mPW1K/6-31+G*          & 3.215  & 1.960  & 1.079  & 108.0  \\
          &  mPW1K/cc-pVDZ(+X)      & 3.304  & 1.968  & 1.086  & 107.6   \\
          &  mPW1K/cc-pVTZ(+X)      & 3.312  & 1.960  & 1.076  & 107.8   \\
          &  MP2/6-31G*             & 3.196  & 1.992  & 1.083  & 107.7   \\
          &  MP2/6-31+G*$^a$    & 3.395  & 1.988  & 1.084  & 107.8   \\
\end{tabular}
$^a$ From Glukhovtsev, M.N.; Pross, A.;  Radom, L.;
{\em J. Am. Chem. Soc.} {\bf 1995}, {\it 117}, 2024.
\end{table}
\newpage

\begin{table}
\caption{\label{y-ch3x}Geometries (\AA\ , degree) of ion-molecule complexes, Y$^-$$\cdots$CH$_3$X,
of the SN$_2$ non-Identity reactions (complex {\bf 1}).}
\squeezetable
\begin{tabular}{llrrrr}
Y/X & Method/Basis Set & r(Y$\cdots$C)   & r(C$-$X) & r(C$-$H) &  $\angle$HCX \\
\hline
F/Cl      &  B3LYP/cc-pVTZ(+X)      & 2.443  & 1.908  & 1.076  &  106.1   \\
          &  BH\&HLYP/cc-pVTZ(+X)  & 2.494  & 1.853 & 1.070 &   107.4 \\
          &  mPW1PW91/cc-pVTZ(+X)   & 2.462  & 1.861  & 1.078  &  107.3   \\
          &  mPWH\&HPW91/cc-pVTZ(+X) & 2.489 &  1.825 & 1.072 & 108.1  \\
          &  B97/cc-pVDZ(+X)        &  2.467  &  1.913  &  1.090  &  106.0  \\
          &  B97-1/cc-pVDZ(+X)      &  2.445  &  1.916  &  1.090  & 105.9  \\
          &  HCTH/cc-pVDZ(+X)       &  2.565  &  1.903  &  1.091  & 106.5 \\
          &  HCTH-120/cc-pVDZ(+X)   &  2.459  &  1.931  &  1.090  & 105.6  \\
          &  mPW1K/6-31+G*          & 2.487  & 1.846  & 1.077  &  107.8  \\
          &  mPW1K/cc-pVDZ(+X)      & 2.495  & 1.837  & 1.083  & 107.8   \\
          &  mPW1K/cc-pVTZ(+X)      & 2.486  & 1.833  & 1.074  &  108.0   \\
          &  MP2/6-31+G*$^a$        & 2.616  & 1.832  &  1.083  & 108.5   \\
          &  B3LYP/6-311+G(2d,d,p)  & 2.410  & 1.931  &  1.078  &  105.5  \\
          &  CCSD(T)/$spdfg$$^b$       &  2.502 & 1.853  &  1.080  &  107.6   \\
\hline
F/Br      &  BH\&HLYP/cc-pVTZ(+X)  & 2.392   & 2.052 & 1.068 & 105.2    \\
          &  mPWH\&HPW91/cc-pVTZ(+X) & 2.404 & 2.007 & 1.070 & 106.4  \\
          &  mPW1K/6-31+G*          & 2.411  & 2.022  & 1.075  &  106.2   \\
          &  mPW1K/cc-pVDZ(+X)      & 2.360  & 2.045  & 1.081  &  105.2   \\
          &  mPW1K/cc-pVTZ(+X)      & 2.389  & 2.022  & 1.072  &  106.0  \\
          &  MP2/6-31+G*$^a$        & 2.528  & 2.028   &  1.081  & 106.8   \\
\hline
Cl/Br     &  B3LYP/cc-pVTZ(+X)      &  3.112  &  2.024  & 1.079   &  106.7   \\
          &  BH\&HLYP/cc-pVTZ(+X)  & 3.149  & 1.986 & 1.072 & 107.3     \\
          &  mPW1PW91/cc-pVTZ(+X)   &  3.088  &  1.989  & 1.080   &  107.4  \\
          &  mPWH\&HPW91/cc-pVTZ(+X) & 3.113 &  1.958 &   1.074 &  107.8  \\
          &  B97/cc-pVDZ(+X)        &  3.134  &  2.021  &  1.093  &  106.8  \\
          &  B97-1/cc-pVDZ(+X)      &  3.098  &  2.022 &  1.093  &  106.8  \\
          &  HCTH/cc-pVDZ(+X)       &  3.269  &  2.012 &  1.094  &  107.1   \\
          &  HCTH-120/cc-pVDZ(+X)   &  3.132  &  2.031  &  1.093  &  106.8  \\
          &  mPW1K/6-31+G*          &  3.070  &  1.966  & 1.079   &  107.9 \\
          &  mPW1K/cc-pVDZ(+X)      &  3.114  &  1.974  & 1.085   &  107.5   \\
          &  mPW1K/cc-pVTZ(+X)      &  3.109  &  1.966 & 1.076    &  107.8 \\
          &  MP2/6-31G*             & 3.092  &  1.992   & 1.082    & 107.6   \\
          &  MP2/6-31+G*$^a$            &  3.199  &  1.992  &  1.084  & 107.7   \\
          &  CCSD(T)/$spdf$$^c$    &   3.095  & 1.986  &  1.082 & 107.5   \\
\end{tabular}
$^a$ From Glukhovtsev, M.N.; Pross, A.;  Radom, L.;
{\em J. Am. Chem. Soc.} {\bf 1996}, {\it 118}, 6273.   \\
$^b$ From  Schmatz, S.; Botschwina, P.; Stoll, H.;
{\em Int. J. Mass Spectrom.} {\bf 2000}, {\it 201}, 277. \\
$^c$ From Botschwina, P.; Horn, M.; Seeger, S.; Oswald, R.; 
{\em Ber. Bunsenges. Phys. Chem.} {\bf 1997}, {\it 101}, 387. \\
\end{table}
\newpage

\begin{table}
\caption{\label{ych3-x}Geometries (\AA\ , degree) of ion-molecule complexes, YCH$_3$$\cdots$X$^-$,
of the SN$_2$ non-identity reactions (complex {\bf 3}).}
\squeezetable
\begin{tabular}{llrrrr}
Y/X & Method/Basis Set & r(X$\cdots$C)   & r(C$-$Y) & r(C$-$H)  & $\angle$HCY \\
\hline
F/Cl      &  B3LYP/cc-pVTZ(+X)      & 3.286  & 1.425  & 1.085  & 108.8   \\
          &  BH\&HLYP/cc-pVTZ(+X)  & 3.268  &  1.401  & 1.077  & 109.0   \\
          &  mPW1PW91/cc-pVTZ(+X)   & 3.246  & 1.409  & 1.085  & 109.2   \\
          &  mPWH\&HPW91/cc-pVTZ(+X) & 3.230 & 1.389 & 1.079  & 109.4  \\
          &  B97/cc-pVDZ(+X)        & 3.320  &  1.432  &  1.098 &  108.7  \\
          &  B97-1/cc-pVDZ(+X)      &  3.268 &  1.432  &  1.098  &  108.7  \\
          &  HCTH/cc-pVDZ(+X)       &  3.528  &  1.431 & 1.099 &  108.7  \\
          &  HCTH-120/cc-pVDZ(+X)   &  3.358  &  1.436  &  1.099  &  108.7  \\
          &  mPW1K/6-31+G*          & 3.241  & 1.403  & 1.083  & 108.8   \\
          &  mPW1K/cc-pVDZ(+X)      & 3.252  & 1.407  & 1.089  & 109.0   \\
          &  mPW1K/cc-pVTZ(+X)      & 3.235  & 1.394  &  1.081  & 109.3   \\
          &  MP2/6-31G*             & 3.227  & 1.415   & 1.087  & 109.4   \\
          &  MP2/6-31+G*$^a$        & 3.255  & 1.438  & 1.086  & 108.0   \\
          &  B3LYP/6-311G(2d,d,p)   & 3.299  & 1.418  & 1.089   & 109.4   \\
          &  B3LYP/6-311+G(2d,d,p)  & 3.271  & 1.430  & 1.087   & 108.7   \\
          &  CCSD(T)/$spdfg$$^b$    &     3.188  & 1.418  &  1.086  &  108.9 \\
\hline

F/Br      &  B3LYP/cc-pVTZ(+X)  &  3.497   &  1.421 & 1.085  &  108.8  \\
          &  BH\&HLYP/cc-pVTZ(+X)  &  3.482 & 1.398 & 1.078 & 109.0  \\ 
          &  mPW1PW91/cc-pVTZ(+X)   & 3.451  &  1.406  &  1.086  &  109.2  \\
           &  mPWH\&HPW91/cc-pVTZ(+X) & 3.433 & 1.386 & 1.080 & 109.3 \\
          &  B97/cc-pVDZ(+X)        & 3.518  &  1.428 &  1.098  &  108.7  \\
          &  B97-1/cc-pVDZ(+X)      &  3.458 &  1.428  &  1.098  &  108.7  \\
          &  HCTH/cc-pVDZ(+X)       &  3.746  &  1.428  &  1.099  &  108.8  \\
          &  HCTH-120/cc-pVDZ(+X)   &  3.565  &  1.433  &  1.100  & 108.7  \\
          &  mPW1K/6-31+G*          & 3.367 & 1.401 & 1.083  & 108.8   \\
          &  mPW1K/cc-pVDZ(+X)      & 3.445 & 1.404  & 1.090  & 109.0   \\
          &  mPW1K/cc-pVTZ(+X)      & 3.443 & 1.391  &  1.081  & 109.3   \\
          &  MP2/6-31G*             & 3.315  & 1.414  &  1.088  & 109.5  \\
          &  MP2/6-31+G*$^a$        & 3.457  & 1.435  & 1.087  & 108.0   \\
\hline
Cl/Br     &  B3LYP/cc-pVTZ(+X)      & 3.405 & 1.839  &  1.081  &  108.0   \\
          &  BH\&HLYP/cc-pVTZ(+X)  & 3.416  &  1.813 & 1.074 & 108.3  \\
          &  mPW1PW91/cc-pVTZ(+X)   & 3.368 & 1.812  &  1.082  &  108.5   \\
           &  mPWH\&HPW91/cc-pVTZ(+X) & 3.369 & 1.791 & 1.076 &  108.7  \\
          &  B97/cc-pVDZ(+X)        & 3.396  &  1.850  &  1.094  &  107.8  \\
          &  B97-1/cc-pVDZ(+X)      & 3.354 &  1.851  &  1.094  &  107.7  \\
          &  HCTH/cc-pVDZ(+X)       & 3.588  &  1.832  &  1.095  &  108.2  \\
          &  HCTH-120/cc-pVDZ(+X)   &  3.440  &  1.842  &  1.095  &  108.0 \\
          &  mPW1K/6-31+G*          & 3.274 & 1.809  &  1.080  &  108.6  \\
          &  mPW1K/cc-pVDZ(+X)      & 3.381 & 1.800  & 1.087   &  108.6   \\
          &  mPW1K/cc-pVTZ(+X)      & 3.371 & 1.797  & 1.077   & 108.7   \\
          &  MP2/6-31G*             & 3.257  & 1.812 & 1.084  & 109.1   \\
          &  MP2/6-31+G*$^a$        & 3.457 & 1.807  & 1.085   &  108.9  \\
          &  CCSD(T)/$spdf$$^c$    &  3.318  &  1.820  &  1.083  &  108.2  \\
\end{tabular}
$^a$ From Glukhovtsev, M.N.; Pross, A.;  Radom, L.;
{\em J. Am. Chem. Soc.} {\bf 1996}, {\it 118}, 6273.  \\
$^b$ From  Schmatz, S.; Botschwina, P.; Stoll, H.;
{\em Int. J. Mass Spectrom.} {\bf 2000}, {\it 201}, 277. \\
$^c$ From Botschwina, P.; Horn, M.; Seeger, S.; Oswald, R.; 
{\em Ber. Bunsenges. Phys. Chem.} {\bf 1997}, {\it 101}, 387. \\
\end{table}
\newpage

\begin{table}
\caption{\label{x-ch3-x}Geometries (\AA\ , degree) of the XCH$_3$X$^-$ transition structures,
(X=F, Cl, and Br) of the SN$_2$ identity reactions (complex {\bf 2}).}
\squeezetable
\begin{tabular}{llrr}
Species & Method/Basis Set & r(X$\cdots$C)    & r(C$-$H)  \\
\hline
X=F       &  CCSD(T)/cc-pVQZ+1      & 1.808  & 1.071    \\
          &  B3LYP/cc-pVTZ(+X)      & 1.854 & 1.070      \\
          &  BH\&HLYP/cc-pVTZ(+X)  &  1.823 & 1.062          \\
          &  mPW1PW91/cc-pVTZ(+X)   & 1.824 & 1.071      \\
           &  mPWH\&HPW91/cc-pVTZ(+X) & 1.797 & 1.064         \\
          &  B97/cc-pVDZ(+X)        & 1.852 & 1.085  \\
          &  B97-1/cc-pVDZ(+X)      & 1.848 &  1.084   \\
          &  HCTH/cc-pVDZ(+X)       & 1.875 &  1.085   \\
          &  HCTH-120/cc-pVDZ(+X)   & 1.872 & 1.086    \\
          &  mPW1K/6-31+G*          & 1.807 & 1.070      \\
          &  mPW1K/cc-pVDZ(+X)      & 1.810 & 1.077      \\
          &  mPW1K/cc-pVTZ(+X)      & 1.804 & 1.066      \\
          &  MP2/6-31G*             & 1.778 & 1.076      \\
          &  MP2/6-31+G*$^a$        & 1.837 & 1.074       \\
          &  B3LYP/6-311G(2d,d,p)   & 1.830 & 1.074      \\
          &  B3LYP/6-311+G(2d,d,p)  & 1.862 & 1.073      \\
\hline                                                        
X=Cl      &  CCSD(T)/cc-pVQZ+1      & 2.305  & 1.070    \\ 
          &  B3LYP/cc-pVTZ(+X)      & 2.355 & 1.069    \\
          &  BH\&HLYP/cc-pVTZ(+X)  &  2.332  & 1.061  \\
          &  mPW1PW91/cc-pVTZ(+X)   & 2.310 & 1.070    \\
       &  mPWH\&HPW91/cc-pVTZ(+X) & 2.290  & 1.064    \\
          &  B97/cc-pVDZ(+X)        & 2.349 & 1.083   \\
          &  B97-1/cc-pVDZ(+X)      & 2.344  &  1.083   \\
          &  HCTH/cc-pVDZ(+X)       & 2.374  & 1.084  \\
          &  HCTH-120/cc-pVDZ(+X)   &  2.372 &  1.084  \\
          &  mPW1K/6-31+G*          & 2.313 & 1.069   \\
          &  mPW1K/cc-pVDZ(+X)      & 2.303 & 1.076   \\
          &  mPW1K/cc-pVTZ(+X)      & 2.295 & 1.065   \\
          &  MP2/6-31G*             & 2.308 & 1.072   \\
          &  MP2/6-31+G*$^a$        & 2.317 & 1.073    \\
          &  B3LYP/6-311G(2d,d,p)   & 2.366 & 1.072   \\
          &  B3LYP/6-311+G(2d,d,p)  & 2.364 & 1.072   \\
\hline                                                        
X=Br      &  CCSD(T)/cc-pVQZ+1     & 2.461  & 1.071   \\
          &  B3LYP/cc-pVTZ(+X)      & 2.511 & 1.069    \\
          &  BH\&HLYP/cc-pVTZ(+X)  & 2.488 & 1.062        \\
          &  mPW1PW91/cc-pVTZ(+X)   & 2.464 & 1.070     \\
          &  mPWH\&HPW91/cc-pVTZ(+X) & 2.442 & 1.064        \\
          &  B97/cc-pVDZ(+X)        & 2.502 &  1.084    \\
          &  B97-1/cc-pVDZ(+X)      &  2.496   &  1.084   \\
          &  HCTH/cc-pVDZ(+X)       &  2.533 &  1.085  \\
          &  HCTH-120/cc-pVDZ(+X)   &  2.530  &  1.085  \\
          &  mPW1K/6-31+G*          & 2.430 & 1.069    \\
          &  mPW1K/cc-pVDZ(+X)      & 2.459 & 1.076    \\
          &  mPW1K/cc-pVTZ(+X)      & 2.447 & 1.066    \\
          &  MP2/6-31G*             & 2.444 & 1.073    \\
          &  MP2/6-31+G*$^a$            & 2.480 & 1.074    \\
\end{tabular}
$^a$ From Glukhovtsev, M.N.; Pross, A.;  Radom, L.;
{\em J. Am. Chem. Soc.} {\bf 1995}, {\it 117}, 2024.
\end{table}
\newpage

\begin{table}
\caption{\label{y-ch3-x}Geometries (\AA\ , degree) of the YCH$_3$X$^-$ transition structures,
of the SN$_2$ non-identity reactions (complex {\bf 2}).}
\squeezetable
\begin{tabular}{llrrrr}
Y/X & Method/Basis Set & r(Y$\cdots$C) & r(C$\cdots$X)  & r(C$-$H) &  $\angle$HCX  \\
\hline
F/Cl    &  B3LYP/cc-pVTZ(+X)      & 2.143 & 2.091 & 1.070  & 99.0     \\
          &  BH\&HLYP/cc-pVTZ(+X)  & 2.065 & 2.107 & 1.062 & 97.2     \\
        &  mPW1PW91/cc-pVTZ(+X)   & 2.072 & 2.086 & 1.071  & 97.8      \\
      &   mPWH\&HPW91/cc-pVTZ(+X) & 2.016 &  2.088 &  1.064 &  96.7    \\
        &  B97/cc-pVDZ(+X)       & 2.143 & 2.091  &  1.085 &  98.8     \\
        &  B97-1/cc-pVDZ(+X)     & 2.139 & 2.086 &  1.085 &  98.8    \\
        &  HCTH/cc-pVDZ(+X)      & 2.190 &  2.098 & 1.086 &  99.2    \\
        &  HCTH-120/cc-pVDZ(+X)  &  2.208 &  2.077 & 1.086 & 100.0    \\
        &  mPW1K/6-31+G*          & 2.063 & 2.080 & 1.070  & 98.0      \\
        &  mPW1K/cc-pVDZ(+X)      & 2.047 & 2.083  & 1.076 & 97.4      \\
        &  mPW1K/cc-pVTZ(+X)      & 2.029 & 2.089 & 1.066  & 97.0     \\
        &  MP2/6-31+G*$^a$        & 2.016 & 2.142 & 1.073  & 95.6     \\
        &  B3LYP/6-311+G(2d,d,p)  & 2.173 & 2.080  &  1.074  & 99.7   \\
          &  CCSD(T)/$spdfg$$^b$    &  2.030  & 2.121  &  1.072  & 96.3   \\
\hline                                                                    
F/Br      &  BH\&HLYP/cc-pVTZ(+X)  & 2.174 & 2.181 & 1.063 &  100.0     \\
          &  mPWH\&HPW91/cc-pVTZ(+X) & 2.105 & 2.175 & 1.065 & 99.2    \\
          &  mPW1K/6-31+G*          & 2.114 & 2.181 & 1.070 & 99.2     \\
          &  mPW1K/cc-pVDZ(+X)      & 2.179 & 2.145 & 1.078 & 100.9    \\
          &  mPW1K/cc-pVTZ(+X)      & 2.129 & 2.168 & 1.067 & 99.7     \\
          &  MP2/6-31+G*$^a$        & 2.108 & 2.242 & 1.075 & 97.9    \\
\hline                                                                   
Cl/Br     &  B3LYP/cc-pVTZ(+X)      & 2.416 & 2.451 & 1.069 & 91.8      \\
          &  BH\&HLYP/cc-pVTZ(+X)  & 2.388  & 2.433 & 1.062 & 91.6      \\
          &  mPW1PW91/cc-pVTZ(+X)   & 2.369 & 2.409 & 1.070 & 91.7     \\
          &  mPWH\&HPW91/cc-pVTZ(+X) & 2.342 & 2.391 & 1.064 &  91.5   \\
          &  B97/cc-pVDZ(+X)        & 2.408  & 2.445 & 1.084 & 91.7  \\
          &  B97-1/cc-pVDZ(+X)      & 2.403  &  2.439 &  1.084  & 91.7  \\
          &  HCTH/cc-pVDZ(+X)       & 2.488 &  2.467  & 1.085 &  91.3  \\
          &  HCTH-120/cc-pVDZ(+X)   & 2.435 &  2.469 &  1.085  & 91.8  \\
          &  mPW1K/6-31+G*          & 2.313 & 2.421 & 1.069 &  90.1     \\
          &  mPW1K/cc-pVDZ(+X)      & 2.363 & 2.401 & 1.076 &  91.8     \\
          &  mPW1K/cc-pVTZ(+X)      & 2.349 & 2.396 & 1.065 &  91.6     \\
          &  MP2/6-31G*             & 2.336 & 2.419 & 1.072 & 90.5      \\
          &  MP2/6-31+G*$^a$        & 2.371 & 2.430 & 1.073 &  91.4       \\
          &  CCSD(T)/$spdf$$^c$    &  2.354  &  2.422  & 1.072   & 91.2  \\
\end{tabular}
$^a$ From Glukhovtsev, M.N.; Pross, A.;  Radom, L.;
{\em J. Am. Chem. Soc.} {\bf 1996}, {\it 118}, 6273.    \\
$^b$ From  Schmatz, S.; Botschwina, P.; Stoll, H.;
{\em Int. J. Mass Spectrom.} {\bf 2000}, {\it 201}, 277. \\
$^c$ From Botschwina, P.; Horn, M.; Seeger, S.; Oswald, R.; 
{\em Ber. Bunsenges. Phys. Chem.} {\bf 1997}, {\it 101}, 387. \\
\end{table}